\documentclass[a4paper,fleqn,usenatbib]{mnras} 

\usepackage{siunitx,booktabs,caption}
\usepackage{newtxtext,newtxmath}
\usepackage[table]{xcolor}

\usepackage{hyperref}

\newcommand{\orcidicon}[1]{\href{https://orcid.org/#1}{\textsuperscript{\includegraphics[width=8pt]{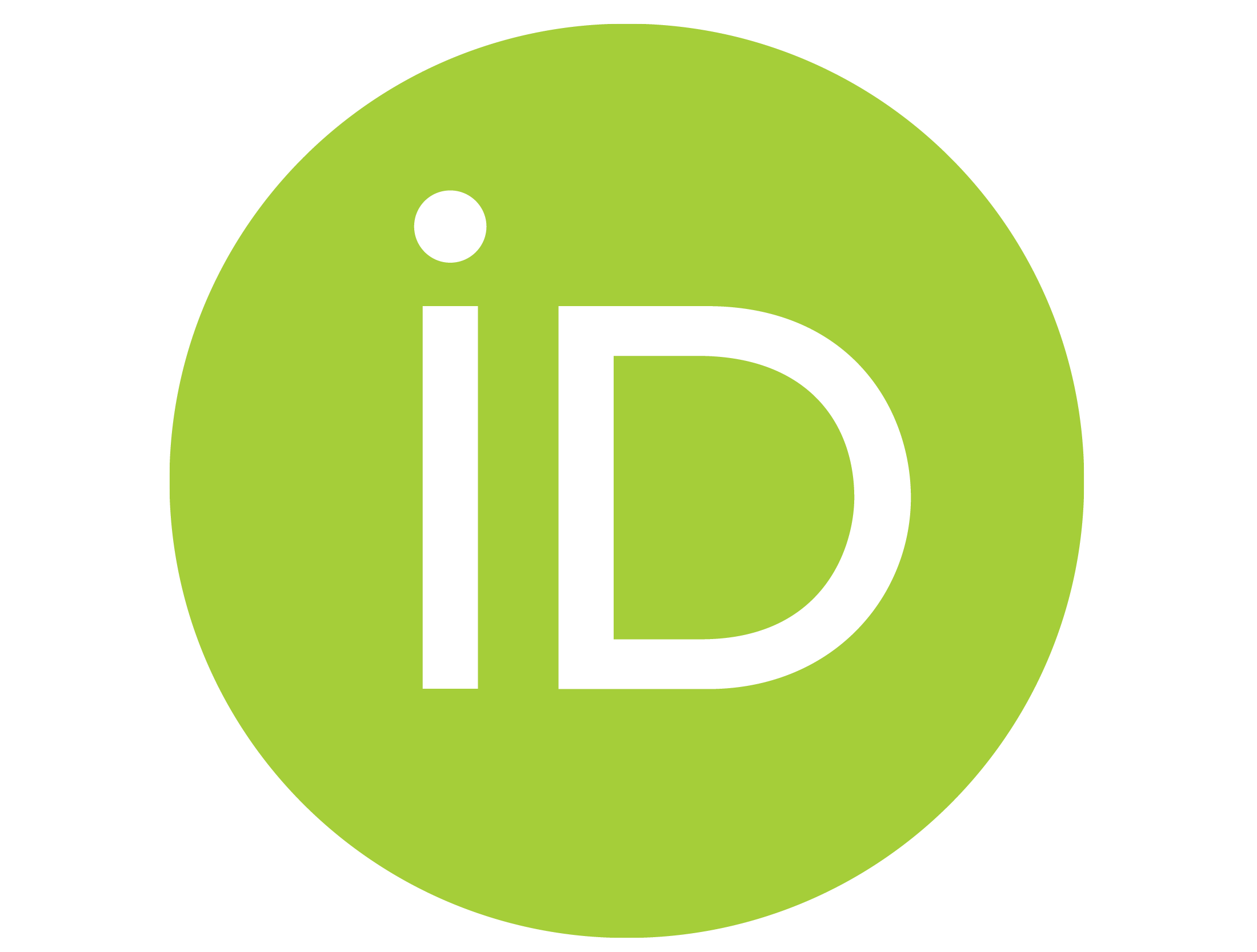}}}}
\usepackage[T1]{fontenc}

\DeclareRobustCommand{\VAN}[3]{#2}
\let\VANthebibliography\thebibliography
\def\thebibliography{\DeclareRobustCommand{\VAN}[3]{##3}\VANthebibliography}

\usepackage{graphicx}	
\graphicspath{{./}{Graphics/}}
\usepackage{amsmath}	
\usepackage{subcaption}

\title[Accretion in Multi-Component Galactic Potentials]{Transonic accretion and the analogue gravity in multi-component elliptical galaxies hosting pseudo-Schwarzschild black holes}

\author[Ripon Sk et al.]{
Ripon Sk \orcidicon{0000-0001-6140-5093},$^{1}$\thanks{riponphysics@gmail.com}
Sangita Chatterjee \orcidicon{0000-0003-3979-113X},$^{2}$ \thanks{sangita6chatterjee@gmail.com}
Sankhasubhra Nag \orcidicon{0000-0001-6373-7040}, $^{3}$ \thanks{sankha@rkmvccrahara.ac.in}
\\
$^{1}$ Department of Physics, West Bengal State University, Berunanpukhuria, West Bengal 700126, India\\
$^{2}$ School of Astrophysics, Presidency University, 86/1 College Street, Kolkata 700073, India.\\
$^{3}$ Department of Physics, Ramakrishna Mission Vivekananda Centenary College, West Bengal 700118, India.\\
}

\date{Accepted XXXX. Received YYYY; in original form ZZZZ}

\pubyear{2025}
\begin{document}
\label{firstpage}
\pagerange{\pageref{firstpage}--\pageref{lastpage}}
\maketitle

\begin{abstract}
Low-angular-momentum, axisymmetric, inviscid accretion flows onto a black hole have been studied using the vertical equilibrium disc
	model, considering multiple pseudo-Schwarzschild potentials and two thermodynamic equations of state. A multi-component
	galactic potential—representing stellar, dark matter, and hot-gas contributions—is incorporated to assess environmental effects
	on the accretion dynamics. In an earlier work \citep{2025arXiv250918833S}, it is found that the effect of multi-component galactic potential on the accretion flow onto a rotating black hole under similar framework of analysis, significantly varies over different standard disc models, being most pronounced in the vertical equilibrium (VE) disc model. Thus it may be interesting to find whether such variation occur for different choices of pseudo potentials too. To begin with, in this work we consider accretion flow onto a non-rotating blackhole with VE  geometry. Through the analysis of transonic behaviour and eigenvalue-based critical point classification, we 	demonstrate that, for all selected black hole potentials, the galactic potential profoundly influences the locations of critical 	points, the shock-allowed parameter space, shock-location, shock-driven flow variables, and acoustic surface gravity.
\end{abstract}





\section{Introduction}
\label{sec:intro}

The Event Horizon Telescope (EHT) observations of M87* constitute a milestone in high-resolution astrophysics, providing the first direct constraints on plasma dynamics at the edge of a supermassive black hole \citep{2019ApJ...875L...1E}. Modelling the millimetre-wave emission with general relativistic magnetohydrodynamic (GRMHD) simulations yields a near-horizon accretion rate of $\dot{M} \simeq (3$–$20)\times 10^{-4},M_{\odot},{\rm yr^{-1}}$ \citep{Akiyama_2021}. These GRMHD frameworks—representing turbulent, magnetically dominated RIAF/MAD–SANE flows with parametrised electron thermodynamics—do not generally include long-lived standing shocks as regulators of $\dot{M}$; instead, uncertainties are dominated by electron heating prescriptions, magnetisation, geometry, and radiative transfer effects \citep{2017MNRAS.470.2367C,Ressler_2020}.

However, recent GRMHD studies of low–angular-momentum accretion have demonstrated the formation and persistence of standing shocks in relativistic flows \citep{2017MNRAS.472.4327S,refId0,2025arXiv250722506M}. Classical and semi-relativistic analyses further show that variations in shock location or stability substantially influence post-shock density and pressure, thereby modulating the local mass inflow \citep{1989ApJ...347..365C,2002ApJ...577..880D,2004ApJ...611..964F,2011ApJ...728..142L,2017MNRAS.472.4327S}. Inward shock migration enhances inflow by compressing the post-shock region, whereas outward displacement promotes expansion and drives outflows \citep{1996ApJ...457..805M,2001ApJ...546..429B,2009ApJ...702..649D}. Although these processes operate locally, they can produce measurable variations in the effective accretion rate near the event horizon.

A further limitation of the existing literature is that most studies treat the black hole as the sole dynamical agent governing accretion, implicitly assuming that the surrounding galactic environment plays no role in shaping the flow \citep{1964ApJ...140..796S,1969Natur.223..690L,1972A&A....21....1P,1973A&A....24..337S,2015NewA...37...81D,2016NewA...43...10S,2017Ap&SS.362...81N,2019PhRvD.100d3024T,2021PhRvD.103b3023T}. In reality, galaxies contain multiple mass components—stellar distribution, hot gas, and a dark matter halo—each contributing to the global gravitational potential. Although weaker than the black hole’s near the horizon, these components can modify the global structure of the accretion solution, particularly in regimes where the flow is sensitive to small perturbations in the effective potential \citep{2018MNRAS.479.3011R,2021JCAP...05..025R}.

Our earlier work \citep{2025arXiv250918833S} incorporated a multi-component galactic potential following \citet{2005MNRAS.363..705M}, consisting of (i) a central Kerr black hole, (ii) a stellar distribution, (iii) a diffuse hot-gas component, and (iv) an extended dark-matter halo. This analysis demonstrated that even modest galactic contributions can induce measurable changes in the flow topology—including systematic shifts in sonic point locations and shock positions—along with corresponding modifications in Mach profiles and stability behaviour. Importantly, variations in shock location are not merely geometric adjustments: they cascade into other physical quantities such as the post-shock density jump, entropy generation, and ultimately the mass accretion rate inferred near the event horizon. These cumulative effects were found to be especially pronounced in the vertical equilibrium (VE) disc model.

Therefore, given that shock formation directly influences several flow variables—including density compression, post-shock temperature, angular momentum redistribution, and consequently the mass accretion rate—it becomes essential to examine how the shock location responds when the accreting gas is subjected not only to the black hole’s gravity but also to the broader galactic potential in more detail considering various previously prescribed black hole potentials. In this context, although general relativity (GR) provides the most rigorous framework for describing accretion near compact objects, its analytical treatment becomes exceedingly complex when extended to systems where gravity arises from multiple mass components \citep{2018MNRAS.479.3011R,2021JCAP...05..025R, 2025arXiv250918833S} . A pragmatic alternative lies in the use of pseudo-Schwarzschild potentials, which approximate the relativistic features of the Schwarzschild metric while preserving the simplicity of Newtonian dynamics. These effective potentials reproduce the strong-field behaviour of black hole gravity---such as the location of marginally stable and bound orbits---without invoking the full machinery of GR. Their use has proven particularly effective for investigating transonic, shock-forming, and advective accretion flows in various astrophysical contexts \citep{1980A&A....88...23P,1996ApJ...461..565A,2001A&A...374.1150D,2002IJMPD..11.1285D,2002IJMPD..11..427S,2002ApJ...577..880D,2003ApJ...588L..89D,2005GReGr..37.1877D,2007arXiv0704.3618D,2007MNRAS.378.1400M,2011PhLB..697..506C,2014CQGra..31c5002B}.

In this present work, we consider four set of pseudo-Schwarzschild potentials, each designed to capture specific relativistic aspects of accretion dynamics around non-rotating black holes. The first and most widely used among them was introduced by \citet{1980A&A....88...23P} have the form,
\begin{equation}
\Phi^{\text{BH}}_1(r) = -\frac{1}{2(r - 1)},
\label{potens1}
\end{equation}
where $r$ is the radial distance presented interms of Schwarzschild radius $r_g=\frac{2GM_{BH}}{c^2}$, $M_{BH}$ represent black hole mass, $G$ and $c$ are the gravitational constant and speed of light. For simplicity we assume $G=M_{BH}=c=1$ throughout the paper. This potential reproduces reproduces the innermost stable circular orbit (ISCO) radius at  $3r_g$ and marginally bound orbit radius at $2r_g$ which is almost same as predicted by GR. Its simplicity and accuracy in describing strong-field effects made it remarkably successful in modelling accretion discs and explaining several observational signatures of X-ray binaries.

The second potential developed by \citet{1991ApJ...378..656N} which had the form,
\begin{equation}
\Phi^{\text{BH}}_2(r) = -\frac{1}{2r}\left[1 - \frac{3}{2r} + 12\left(\frac{1}{2r}\right)^2\right],
\label{potens2}
\end{equation}
In their study of acoustic oscillation normal modes in thin accretion discs around compact objects—specifically slowly rotating black holes or weakly magnetized neutron stars—Nowak and Wagoner (1991) modeled the leading relativistic corrections by introducing this modified Newtonian potential.
This potential exhibits smoother behaviour near the horizon and reproduces relativistic epicyclic frequencies with reasonable accuracy, making it suitable for studying high-frequency quasi-periodic oscillations (QPOs).

Both the third and the fourth potential were proposed by \citet{1996ApJ...461..565A}. The third potential reproduces the Schwarzschild coordinate free-fall acceleration of a particle that is momentarily at rest at radius $r$ and the fourth logarithmic potential defined so that its acceleration reproduces the covariant radial component of the Schwarzschild three-dimensional free-fall acceleration vector for a particle at rest.
\begin{equation}
\Phi^{\text{BH}}_3(r) = -1 + \sqrt{1 - \frac{1}{r}},
\label{potens3}
\end{equation}

\begin{equation}
\Phi^{\text{BH}}_4(r) = \frac{1}{2}\ln\left(1 - \frac{1}{r}\right),
\label{potens4}
\end{equation}

Among the two potentials, $\Phi^{\text{BH}}_3(r)$ is slightly more efficient because it reproduces the Schwarzschild free-fall acceleration while remaining numerically smooth for transonic and shock-forming flows. $\Phi^{\text{BH}}_4(r)$ matches the covariant GR acceleration more closely, but its steeper near-horizon behaviour often makes it less stable in practical simulations \citep{2002ApJ...577..880D}.

From this point onward, we shall collectively refer to all four of these potentials as $\Phi^{\text{BH}}_i(r)$, where $i = 1, 2, 3, 4$ corresponds respectively to $\Phi^{\text{BH}}_1(r)$ [Eq.~(\ref{potens1})], $\Phi^{\text{BH}}_2(r)$ [Eq.~(\ref{potens2})], $\Phi^{\text{BH}}_3(r)$ [Eq.~(\ref{potens3})], and $\Phi^{\text{BH}}_4(r)$ [Eq.~(\ref{potens4})]. A comparative overview of the physical characteristics and applicability of these potentials has been presented in \citet{2001A&A...374.1150D}, \citet{2002ApJ...577..880D}, and \citet{2003ApJ...592.1078D}. Further discussions can be found in \citet{2006MNRAS.373..146C} and \citet{Bilić2014}.

In the present work, we embed each of the pseudo-Schwarzschild black hole potentials, $\Phi^{\text{BH}}_i(r)$, within a realistic galactic potential characterized by the parameter $\Upsilon_B$, which encapsulates the combined gravitational influence of the stellar, dark matter, and gaseous components. Following our previous work \citet{2025arXiv250918833S} we presented the stellar mass distribution by a Sérsic profile \citep{1968adga.book.....S}, the diffuse hot gas is represented by an isothermal $\beta$-model \citep{2001ApJ...547..154B,2005MNRAS.363..705M}, and for the dark matter halo we adopts the generalized NFW-type profile of \citet{2000ApJ...529L..69J}. We use the polyropic equation of state and vertical profile of the disc is presented by vertical equilibrium disc model.

Our study shows that the global transonic structure of low–angular-momentum, inviscid, axisymmetric accretion flows onto a Schwarzschild black hole undergoes substantial modification when a realistic multi-component galactic potential is incorporated. The inclusion of galaxy potential noticeably shifts the positions of the critical points—particularly the outer saddle—by altering the effective gravitational gradient experienced by the flow. As a result, both the number and radial distribution of sonic points change across all four pseudo-Schwarzschild potentials considered under the vertical equilibrium disc model. This modification of the sonic framework directly influences the subsequent formation and evolution of standing shocks. Once the critical structure is altered, the Rankine–Hugoniot shock location responds sensitively and exhibits systematic relocation in the presence of the galactic potential. Our results show that increasing the galactic parameter $\Upsilon_B$ compresses the multitransonic region in parameter space, shrinks and displaces the shock-permitting domain, and modifies the compression ratio and shock strength accordingly. These dynamical changes propagate further into the analogue gravity sector, where the acoustic surface gravity $\kappa$ exhibits systematic variation with $\Upsilon_B$, reflecting the dilution of the effective gravitational field near the sonic surface. Altogether, the comparative behaviour across the four potentials demonstrates that realistic galactic environments not only reshape the critical-point topology and shock dynamics but also influence many observable properties of accretion flows, emphasizing the necessity of accounting for galactic-scale effects in accurate accretion modelling.

We proceed through the paper in the following steps:: in \S\ref{galpotsec}, we describe the combined gravitational contribution to the potential of the multi-component galaxy. In \S\ref{flowD}, we describe the hydrodynamics of the flow. The nature of the fixed points are discussed in \S\ref{fixednature}. The shock conditions for adiabatic and isothermal flows are discussed in \S\ref{shok-F}. The time depenedent stability of the solutions are described in \S\ref{stable-F}. We expressed the acoustic surface garvity in\S\ref{Surgrav}. Finally the result and discussion are given in \S\ref{Result}. The concluding remarks are given in \S\ref{conclu}.

\section{Model Equations}

\subsection{Gravitational Potential of the elliptical galaxy}
\label{galpotsec}
Elliptical galaxies, typically approximated as nearly spherical, enable simplified modeling of gravitational dynamics. Observationally, their mass distributions exhibit minimal deviation from spherical symmetry, particularly in massive systems. The gravitational force arising from a spherically symmetric mass distribution—in the case of an elliptical galaxy—results from the combined contribution of mainly four mass components: the central black hole (BH), the stellar content, dark matter (DM), and the surrounding diffuse hot gas. The total gravitational potential can thus be represented as a straightforward linear sum of the individual gravitational potential contributions from each of its distinct components, and can be written as,
\begin{equation}
 \Phi^{\text{Gal}}_i(r) =  \Phi^{\text{BH}}_i(r) + \Phi_{\mathrm{star}}(r) + \Phi_{\mathrm{gas}}(r) + \Phi_{\mathrm{DM}}(r)
\label{eqgalphi}
\end{equation}
where $\Phi^{\text{Gal}}_i(r)$ provides the galactic potential and the respective force function can be obtained as $\mathbb{F}_{\mathrm{Gal}}(r)=\frac{d \Phi^{\text{Gal}}_i(r)}{dr}$.

Recent advances in the photometric analysis of elliptical galaxies have shown that traditional models—such as the Hubble–Reynolds law \citep{1913MNRAS..74..132R}, King models \citep{1962AJ.....67..471K}, and the de Vaucouleurs $ R^{1/4}$ law \citep{1948AnAp...11..247D} —generally fail to accurately fit their surface brightness profiles. However, a broad consensus has emerged supporting the Sérsic law, a generalization of the $R^{1/4}$ profile, as a superior and widely applicable model for describing the surface photometry of nearly all elliptical galaxies. The $R^{1/4}$ model proposed by Sérsic \citep{1968adga.book.....S} is most commonly expressed in the form of an intensity profile : $I(r)=I_0 \exp{\left[-(r/r_s)^{1/n}\right]}$ where $I_0$ is the intensity at the effective radius $r_s$ also known as Sérsic scale radius, $n$ is the Sérsic index. Following \citet{2005MNRAS.362...95M,2005MNRAS.363..705M,2005MNRAS.362..197T,2005PASA...22..118G,2018MNRAS.479.3011R}, we get the gravitational potential for the stellar mass distribution in the elliptical galaxy as,

\begin{equation}
\Phi_{\mathrm{star}}= -\int \mathbb{F}_{\mathrm{star}}(r)\,dr = -\int \frac{G f_{\mathrm{star}} M_{v}}{r^2}
   \cdot \frac{ \xi\left[(3{-}\mu)n,\left(\tfrac{r}{r_s}\right)^{1/n}\right] }
              { \xi\left[(3{-}\mu)n,\left(\tfrac{r_v}{r_s}\right)^{1/n}\right] } \, dr
\label{eqstar}
\end{equation}
where $r_{v}\simeq 206.262 h_{70}^{-1}\left(\frac{h_{70}M_{v}}{10^{12}M_{\odot}}\right)^{\frac{1}{3}}$ kpc is the virial radius, and within this virial radius $M_v$ is mass \& $f_{star}$ is stellar mass fraction. The relation between the Sérsic index and $\mu$ is $\mu(n)\simeq 1.0-\frac{0.06097}{n}+\frac{0.05463}{n^2}$. The standard incomplete gamma function is denoted here by, $\xi(t,x)$. The rest of the notations and their details are defined in \citet{1997A&A...321..111P,2005MNRAS.363..705M,2018MNRAS.479.3011R}.

In elliptical galaxies, hot diffuse gas contributes significantly to both X-ray emission and the total baryonic mass, especially at the virial radius. Assuming the cosmic baryon fraction applies, the mass in hot gas may exceed that in stars by a factor of $\sim 4$ \citet{2005MNRAS.362...95M,2005MNRAS.363..705M}. This dominance depends on the galaxy's mass-to-light ratio and suggests that hot gas is a crucial component of the baryon budget in massive ellipticals. Assuming hydrostatic equilibrium with the galaxy’s gravitational potential, the spatial distribution of this hot gas is effectively described by the isothermal $\beta$-model \citep{1976A&A....49..137C,RevModPhys.58.1,2001ApJ...547..154B,2005MNRAS.362...95M,2005MNRAS.363..705M}, which assumes constant temperature and radially varying density, reproducing observed X-ray surface brightness profiles. We employ this isothermal $\beta$-model to represent the hot gas component in our analysis. The associated potential profile is obtained by evaluating the integration of the gravitational force arising from this gas distribution within the virial radius is expressed as:

\begin{equation}
\Phi_{\mathrm{gas}}= -\int \mathbb{F}_{\mathrm{gas}}(r)\,dr = -\int \frac{G f_{\mathrm{gas}} M_\nu}{r^2} \cdot \frac{\mathbb{F}_{N}}{\mathbb{F}_D}.dr \label{eqgas}
\end{equation}
where, $\mathbb{F}_{N}=\left[\left(\frac{1}{3}\left(\frac{r}{r_b}\right)^3\right)^{-\delta} + \left(\frac{2}{3}\left(\frac{r}{r_b}\right)^{3/2}\right)^{-\delta}\right]^{-1/\delta}$ and $\mathbb{F}_D=\left[\left(\frac{1}{3}\left(\frac{r_\nu}{r_b}\right)^3\right)^{-\delta} + \left(\frac{2}{3}\left(\frac{r_\nu}{r_b}\right)^{3/2}\right)^{-\delta}\right]^{-1/\delta}$. Within the virial radius the hot gas mass fraction is given by $f_{gas}$ and scale radius $r_b\simeq \mathcal{R}_{eff}/q$, here $\mathcal{R}_{eff}$ is the effective radius, $q\simeq10$ and $\delta=2^{\frac{1}{8}}$.

Constraints on dark matter in elliptical galaxies are significantly improved by combining internal stellar kinematics with either X-ray observations or strong gravitational lensing. On the theoretical front, high-resolution dissipationless cosmological N-body simulations have converged on halo structures characterized by an outer density slope of $-3$ and inner slopes ranging from $-1$ (NFW) to $-1.5$\citep{1999MNRAS.310.1147M}. \citet{2000ApJ...529L..69J} proposed a generalized double power-law profile that accurately reproduces these simulated halo properties which is given by,
\begin{equation}
\rho_{\mathrm{DM}}(r) \propto \frac{1}{\left(r/r_d\right)^\gamma \left[1+\left(r/r_d\right)\right]^{3-\gamma}}, \quad \gamma = \frac{3}{2}~(\text{JS-1.5})  \nonumber
\end{equation}
with scale radius $r_d = r_{\nu}/c_0$, where $c_0$ is the concentration parameter \citep{2006AJ....132.2685M}.
The gravitational potential associated with this JS-1.5 model's dark matter mass distribution is given by,
\begin{multline}
 \Phi_{\mathrm{DM}} = -\int \mathbb{F}_{\mathrm{DM}}(r)\,dr \\ = -\int \frac{G f_{DM}M_{\nu}}{r^2} \frac{1}{\mathcal{G}(c_0)}2[sinh^{-1}(\sqrt{r/r_d})-\sqrt{\frac{r/r_d}{(1+r/r_d)}}].dr \label{eqdm}
 \end{multline}
Now using these equations in equation \eqref{eqgalphi} we get the host galaxy's gravitational potential $\Phi^{\text{Gal}}_i$.
%
%

For more details see \cite{2025arXiv250918833S}. Using equations \eqref{eqstar}-\eqref{eqdm} in equation \eqref{eqgalphi} we get the multi-component galactic potential $\Phi_{\mathrm{Gal}}$. The main parameters determining the galactic potential of an elliptical galaxy are summarized in Appendix~\ref{galpotd} where we define the four chosen cases of $\Upsilon_B$ which we use to obtain the impact of galctic contribution on the accretion dynamics. For illustration, Fig.~\ref{ellpgal} shows the mass distributions for four representative models stated above. From the figure we can see that for for $\Upsilon_B=14$, the hot gas dominates, for $\Upsilon_B=33$ stars and gas are comparable, while for $\Upsilon_B=100$ and $\Upsilon_B=390$ dark matter is the primary contributor. These four cases are used to explore the effect of  $\Upsilon_B$  on the full accretion dynamics.

\subsection{Galactic Potential Parameters and the Role of $\Upsilon_B$}
\label{galpotd}
To compute the galactic potential function $\Phi_{\mathrm{Gal}}$ of an elliptical galaxy, the key parameter is the \emph{B-band mass-to-light ratio}
\begin{equation}
\Upsilon_B \equiv \frac{M_{\mathrm{tot}}}{L_B},
\end{equation}
which quantifies the total virial mass $M_{\mathrm{tot}}=M_v$ (including stars, gas, dark matter, and black hole) per unit of blue-band luminosity $L_B$. This ratio provides a natural measure of whether a galaxy is baryon-dominated or dark-matter-dominated. The virial mass is related to the luminosity through ~\cite{2015IJMPD..2450084G,2010MNRAS.407.1148C}
\begin{equation}
M_v = b_{\Upsilon}\,\bar{\Upsilon}_B\,L_B ,
\end{equation}
where $\bar{\Upsilon}_B \simeq 390\,h_{70}\,M_\odot/L_\odot$ is the cosmic mean B-band mass-to-light ratio and $b_\Upsilon=\Upsilon_B/\bar{\Upsilon}_B$ is the mass-to-light bias factor.

The stellar contribution is described by the stellar mass-to-light ratio in the B-band, taken as $\Upsilon_{\ast B} \simeq 6.5$, giving the stellar mass fraction ~\cite{2006MNRAS.370.1581M,2005MNRAS.363..705M}
\begin{equation}
f_{\ast} = \frac{\Upsilon_{\ast B}}{\Upsilon_B}.
\end{equation}
The central black hole mass is estimated from the bulge mass $M_{\text{bulge}}=f_{\ast}M_v$ using the empirical correlation  ~\cite{2000ApJ...539L..13G,2002ApJ...574..740T,2013ApJ...764..151G,2004ApJ...604L..89H}
\begin{equation}
\log \!\left(\frac{M_{\mathrm{BH}}}{M_\odot}\right) = (8.20 \pm 0.10) + (1.12 \pm 0.06)\,\log \!\left(\frac{M_{\text{bulge}}}{10^{11} M_\odot}\right).
\end{equation}

The baryon fraction within the virial radius is written as $f_b$, and its deviation from the universal mean $\bar{f}_b=\Omega_b/\Omega_m\simeq0.14$ is quantified by the baryon fraction bias, $b_b=f_b/\bar f_b$. The dark matter and gas mass fractions then follow as ~\cite{2005MNRAS.363..705M,2015IJMPD..2450084G,2018MNRAS.479.3011R}
\begin{equation}
f_{\mathrm{DM}} = 1 - b_b\bar f_b , \qquad
f_{\mathrm{gas}} = 1 - (f_{\ast} + f_{\mathrm{DM}} + f_{\mathrm{BH}}),
\end{equation}
with the black hole fraction $f_{\mathrm{BH}}$ being negligible compared to stars and gas.

In this study we adopt a fiducial luminosity $L_B \simeq 2 \times 10^{10}L_\odot$ and examine four representative values of the global M/L ratio, $\Upsilon_B=14, 33, 100,$ and $390$, which span the full range from baryon-dominated to dark-matter-dominated systems. The lowest values ($\sim 14{-}17$) correspond to galaxies with little or no dark matter, such as NGC~821, while the highest value ($\Upsilon_B\simeq 390$)~\cite{2003Sci...301.1696R} reflects the cosmic mean, consistent with virialized halos. Intermediate cases ($\Upsilon_B=33$ and $100$) describe systems with progressively larger dark matter fractions, such as NGC~3379. For the case $\Upsilon_B=100$, we assume $f_b=\bar f_b\simeq0.14$, while for $\Upsilon_B=33$ a similar gas-to-star ratio is adopted.

The resulting fractions are:
\begin{itemize}
    \item \textbf{Set 1 ($\Upsilon_B=390$)}: $b_\Upsilon=1.0$, $f_\ast=0.0167$, $b_b=1.0$, $f_{\mathrm{DM}}=0.860$, $f_{\mathrm{gas}}=0.123$.\\
    Galaxy Type : Dark-matter-dominated, faint stellar content (e.g.,cluster-central ellipticals such as M87 (NGC 4486) in Virgo ~\cite{1990AJ.....99.1823M})
    \item \textbf{Set 2 ($\Upsilon_B=100$)}: $b_\Upsilon=0.256$, $f_\ast=0.065$, $b_b=1.0$, $f_{\mathrm{DM}}=0.860$, $f_{\mathrm{gas}}=0.075$.\\
    Galaxy Type : Intermediate, still DM-dominated (e.g.,NGC 4494~\cite{2003Sci...301.1696R})
    \item \textbf{Set 3 ($\Upsilon_B=33$)}: $b_\Upsilon=0.085$, $f_\ast=0.197$, $b_b=3.03$, $f_{\mathrm{DM}}=0.576$, $f_{\mathrm{gas}}=0.227$.\\
    Galaxy Type : Mixed, significant stars and gas (e.g.,NGC 3379~\cite{2003Sci...301.1696R})
    \item \textbf{Set 4 ($\Upsilon_B=14$)}: $b_\Upsilon=0.036$, $f_\ast=0.464$, $b_b=7.14$, $f_{\mathrm{DM}}=0.000$, $f_{\mathrm{gas}}=0.536$.\\
    Galaxy Type : Baryon-rich, gas-dominated, negligible DM (e.g.,NGC 821 ~\cite{2003Sci...301.1696R})
\end{itemize}

These sets clearly demonstrate the role of $\Upsilon_B$: low values correspond to luminous, baryon-rich galaxies with strong stellar and gaseous components, while high values describe dark-matter-dominated systems with faint stellar populations. For more details regarding the galactic parameters and other significant parameters please see ~\cite{2005MNRAS.362...95M,2005MNRAS.363..705M,2018MNRAS.479.3011R}.

In the next section, we describe the axisymmetric flow under the impact of the galactic potential  $\Phi^{\text{Gal}}_i$ for adiabatic and isothermal thermodynamic equation of state.

\subsection{Hydrodynamic Equations : Stationary solutions and fixed points  }
\label{flowD}
\subsubsection{Adiabatic flows}
For a steady axisymmetric inviscid flow, in the framework of a  black hole, we can write the Euler equation as \citep{2002apa..book.....F},
\begin{equation}
v \frac{\mathrm{d} v}{\mathrm{d} r} + \frac{1}{\rho} \frac{\mathrm{d} P}{\mathrm{d} r} + \overset{\boldsymbol{\cdot}}{\Phi}{}^{\text{Gal}}_{i}(r) - \frac{\lambda^{2}}{r^{3}} = 0.
\label{euler2}
\end{equation}
and the mass conservation (continuity) equation as,
\begin{equation}
\frac{\mathrm{d}}{\mathrm{d} r} \left( \rho v r H \right) = 0.
\label{cont2}
\end{equation}
where \(\rho\) is the mass density, \(v\) is the radial velocity. \(\lambda\) is the conserved specific angular momentum, and $\overset{\boldsymbol{\cdot}}{\Phi}^{\text{Gal}}_i(r) = \frac{d\Phi^{\text{Gal}}_i(r)}{dr}$ is the spatial derivative with respect to radial distance $r$ and obtained from equation \eqref{eqgalphi}. The pressure \(P(\rho)\) is related to the mass density by the equation of state:
\begin{equation}
P = K \rho^\gamma
\label{adia1}
\end{equation}
where  \( \gamma \) is the adiabatic index, and \( K \) is a constant related to the entropy of the flow \citep{1939isss.book.....C}. Using this equation of state , the specific energy equation also called the Bernoulli equation can be obtained from the first integral of equation \eqref{euler2},
\begin{equation}
\mathcal{E} = \frac{v^{2}}{2} + \frac{c_{s}^{2}}{\gamma - 1} + \frac{\lambda^{2}}{2r^{2}} +  \Phi^{\text{Gal}}_i(r)
\label{ber}
\end{equation}
where \(c_s = \sqrt{\frac{\partial P}{\partial \rho}}\) is the local adiabatic sound speed. Another constant quantity obtained from the first integral of the continuity equation and using the connection of $P$ and $\rho$ is known as the entropy accretion rate \(\dot{\mathcal{M}}\) \citep{1990ttaf.book.....C} that measures the inward flux of entropy per unit time, given by,

\begin{equation}
\dot{\mathcal{M}}^{2} = 4 \pi^{2} c_{s}^{4n} v^{2} r^{2} H^{2} \label{ent}
\end{equation}
where $\dot{\mathcal{M}} = (\gamma K)^{n} \dot{m}$, $n = (\gamma - 1)^{-1}$.
\(\dot{m}\) is the mass accretion rate and $H$ denotes the height of the accretion disc.

In astrophysics, especially when studying accretion discs around compact objects, or active galactic nuclei, the disc height denotes the vertical scale or geometrical thickness of the disc as a function of radial distance from the central object. This height can be described in different ways, depending on the physical assumptions and the dynamical regime of the accretion flow. When the vertical thickness of the disc is much smaller than its radial extent, and the outflow rate is insignificant compared to the total accretion rate, the disc can be effectively described using the vertical equilibrium approximation (VE)—a reliable method for modeling the three-dimensional structure of the disc. In this model, the disc height results from enforcing hydrostatic balance along the vertical direction and is expressed as
\begin{equation}
H(r) = c_{\mathrm{s}} \left( \frac{r}{\gamma \, \overset{\boldsymbol{\cdot}}{\Phi}{}^{\text{Gal}}_{i}(r)} \right)^{1/2},
\label{height1}
\end{equation}

To locate the critical points of the flow, we differentiate  Eqs. \eqref{ber} and \eqref{ent} and then add them to obtain,

\begin{equation}
\frac{d v^{2}}{dr}
=
\frac{\mathcal{N}}{\mathcal{D}}
\label{cri1}
\end{equation}

\begin{equation}
\mathcal{N}
=
2 v^{2}
\left(
\frac{\lambda^{2}}{r^{2}}
- r\,\dot{\Phi}^{\mathrm{Gal}}_{i}(r)
+ c_{\mathrm{s}}^{2}
\left[1 + r \frac{\dot H}{H}\right]
\right)
\end{equation}

\begin{equation}
\mathcal{D}
=
r\left(v^{2}-c_{\mathrm{s}}^{2}\right)
\end{equation}

then using (\ref{height1}), we can simply as
\begin{align}
\left.\frac{dv}{dr}\right|_{\mathrm{}}^{\mathrm{adia}}
&=
\frac{\mathcal{N}_1}{\mathcal{D}_1},
\label{dvdr_nd}
\\
\mathcal{N}_1
&=
v \left[
\frac{c_{s}^{2}}{1 + \gamma}
\left(
\frac{3}{r}
-
\frac{\ddot{\Phi}^{\mathrm{Gal}}_{i}(r)}
{\dot{\Phi}^{\mathrm{Gal}}_{i}(r)}
\right)
+
\frac{\lambda^{2}}{r^{3}}
-
\dot{\Phi}^{\mathrm{Gal}}_{i}(r)
\right],
\\
\mathcal{D}_1
&=
v^{2}
-
\frac{2}{1 + \gamma} \, c_{s}^{2},
\\
\left.\frac{dc_s}{dr}\right|_{\mathrm{}}^{\mathrm{adia}}
&=
\frac{\mathcal{N}_2}{\mathcal{D}_2},
\label{dcsdr_nd}
\\
\mathcal{N}_2
&=
(1 - \gamma)\, c_{s}
\left[
\frac{dv}{dr}
+
\frac{v}{2}
\left(
\frac{3}{r}
-
\frac{\ddot{\Phi}^{\mathrm{Gal}}_{i}(r)}
{\dot{\Phi}^{\mathrm{Gal}}_{i}(r)}
\right)
\right],
\\
\mathcal{D}_2
&=
(1 + \gamma)\, v.
\end{align}

 The critical points of the flow are determined by using the condition that  \(\mathrm{d}v/\mathrm{d}r=0\). Explicitly expressing and rearranging the terms, we obtain the two critical point conditions as:

\begin{equation}
v_{r_c}
=
\left( \frac{2}{\gamma + 1} \right)^{1/2}
c_{s_{r_c}},
\qquad
c_{s_{r_c}}^{2}
=
(\gamma + 1)
\frac{
\dot{\Phi}^{\mathrm{Gal}}_{i}(r_c)
-
\frac{\lambda^{2}}{r_c^{3}}
}{
\frac{3}{r_c}
-
\frac{\ddot{\Phi}^{\mathrm{Gal}}_{i}(r_c)}
{\dot{\Phi}^{\mathrm{Gal}}_{i}(r_c)}
}.
\label{vc}
\end{equation}
For a given set of fixed values of \([\mathcal{E}, \lambda, \gamma]\), the location of the critical point for a specific flow model can be determined by substituting the relevant critical point condition (as described in Eqs. (\ref{vc}) into the energy first integral  (\ref{ber}) for the galactic potential using \citep{2006MNRAS.373..146C,Bilić2014,2016NewA...43...10S}..
\begin{align}
\mathcal{E}
&=
\frac{2\gamma}{\gamma - 1}
\frac{\mathcal{A}_1}{\mathcal{B}_1}
-
\Phi^{\mathrm{Gal}}_{i}(r_c)
-
\frac{\lambda^{2}}{2 r_c^{2}},
\label{E-space}
\\[8pt]
\text{where}\qquad
\mathcal{A}_1
&=
r_c \dot{\Phi}^{\mathrm{Gal}}_{i}(r_c)
-
\frac{\lambda^{2}}{r_c^{2}},
\\[6pt]
\mathcal{B}_1
&=
3
-
r_c
\left(
\frac{\ddot{\Phi}^{\mathrm{Gal}}_{i}}
{\dot{\Phi}^{\mathrm{Gal}}_{i}}
\right)_{r_c}.
\end{align}
From this expression, it is evident that the solutions for \(r_c\) can also be obtained as a function of \(\lambda\) and \(\mathcal{E}\), i.e., \(r_c = f_1(\lambda, \mathcal{E})\). After substitution, the energy first integral transforms into an algebraic equation in terms of \(r_c\). Alternatively, \(r_c\) can be expressed in terms of \(\lambda\) and \(\dot{\mathcal{M}}\). By making use of the critical point conditions in Eq.\eqref{ent}, one can write:
\begin{align}
\dot{\mathcal{M}}^{2}
&=
\frac{
8\pi^{2} r_c^{3}
}{
\gamma(\gamma + 1)\,
\dot{\Phi}^{\mathrm{Gal}}_{i}(r_c)
}
\left[
\frac{\mathcal{A}_2}
     {(\gamma + 1)\,\mathcal{B}_2}
\right]^{2(n+1)},
\label{dotmfix}
\\[8pt]
\text{with}\qquad
\mathcal{A}_2
&=
r_c \dot{\Phi}^{\mathrm{Gal}}_{i}(r_c)
-
\frac{\lambda^{2}}{r_c^{2}},
\\[6pt]
\mathcal{B}_2
&=
3
-
r_c
\frac{\ddot{\Phi}^{\mathrm{Gal}}_{i}(r_c)}
     {\dot{\Phi}^{\mathrm{Gal}}_{i}(r_c)}.
\end{align}
with the critical radius as a function \(r_c = f_2(\lambda, \dot{\mathcal{M}})\), with the obvious implication that the location of the critical point depends explicitly on both angular momentum and entropy accretion rate.

\subsubsection{Isothermal flows}
An isothermal flow refers to a fluid flow in which the temperature remains constant, and the pressure $P$ and density $\rho$ are related by :

\begin{equation}
 P=\frac{\rho \kappa_B T}{\mu m_{H}}
\end{equation}
where, $\kappa_B$, $\mu$, and $m_H$ defines the Boltzmann’s constant, reduced mass \& the mass of a hydrogen atom respectively. $T$ is the constant temperature. For isothermal flow, the integral solution of the time independent Euler equation provides the following first integral of motion
\begin{equation}
\frac{v^2}{2} + c_{\mathrm s}^2 \ln \rho
+ \frac{\lambda^2}{2 r^2} + \Phi^{\text{Gal}}_i(r) = {\rm C}
\label{iso1}
\end{equation}
Obviously, this $C=$ constant of motion can not be identified with the
specific energy of the flow. The mass accretion rate, another first integral of motion of the accreting system of aforementioned kind, may be obtained for three different flow geometries as
\begin{align}
\dot{\mathcal{M}}_{\rm } &= c_s \rho v r^{3/2} \left( \overset{\boldsymbol{\cdot}}{\Phi}{}^{\text{Gal}}_{i}(r) \right)^{-1/2}
\end{align}

The space gradient of the velocities for these three models comes out to be

\begin{align}
\left.\frac{dv}{dr}\right|_{\mathrm{}}^{\mathrm{iso}}
&=
\frac{\mathcal{N}_{\mathrm{iso}}}
     {\mathcal{D}_{\mathrm{iso}}},
\label{dvdr_iso_nd}
\\[8pt]
\mathcal{N}_{\mathrm{iso}}
&=
v \left[
\frac{c_s^{2}}{2}
\left(
\frac{3}{r}
-
\frac{\ddot{\Phi}^{\mathrm{Gal}}_{i}(r)}
     {\dot{\Phi}^{\mathrm{Gal}}_{i}(r)}
\right)
-
\dot{\Phi}^{\mathrm{Gal}}_{i}(r)
+
\frac{\lambda^{2}}{r^{3}}
\right],
\\[8pt]
\mathcal{D}_{\mathrm{iso}}
&=
v^{2} - c_s^{2}.
\end{align}
which provide the following critical point conditions:

\begin{align}
v_{r_c}
&=
c_{s_{r_c}}
=
\left( \frac{\kappa_B}{\mu m_H} \right)^{1/2} T^{1/2}
=
\left[
\frac{2\,\mathcal{A}_3}
     {\mathcal{B}_3}
\right]^{1/2},
\label{iso_rc_AB}
\\[8pt]
\text{where}\qquad
\mathcal{A}_3
&=
r_c \dot{\Phi}^{\mathrm{Gal}}_{i}(r_c)
-
\frac{\lambda^{2}}{r_c^{2}},
\qquad
\mathcal{B}_3
=
3
-
r_c
\frac{\ddot{\Phi}^{\mathrm{Gal}}_{i}(r_c)}
     {\dot{\Phi}^{\mathrm{Gal}}_{i}(r_c)}.
\end{align}

More detailed Discussions and derivations of all the above mentioned equations can be found in \citet{2006MNRAS.373..146C,Bilić2014,2016NewA...43...10S}.
Starting with the critical values of the velocity and its radial derivative, we numerically integrate the governing equations to examine the flow trajectories and phase protraits. Depending on the chosen boundary conditions and flow parameters, this method offers a quantitative representation of the possible flow configurations, either smooth or containing shocks (see figure~\ref{PP1}). However, the qualitative behaviour of the accretion flow cannot be well described by such topological diagrams alone.  Although a trajectory is always produced by numerical integration, it does not reveal if its corresponding critical point is physically achievable.  To determine the true qualitative nature of the flow, a complete dynamical systems analysis has been performed in \S~\ref{fixednature}, taking into account the influence of the multi-component galactic potential.

In the subsequent section, we examine the characteristics of the critical points obtained in the presence of this composite galactic potential.

\subsection{Nature of The Fixed Points: Categorization of Critical Points}
\label{fixednature}
The governing fluid equations that characterise the accretion flow are fundamentally first-order nonlinear differential equations. Due of their nonlinear characteristics, obtaining exact analytical solutions is generally impractical.Therefore, in most realistic situations, , it is necessary to use numerical approaches to integration to investigate both the behaviour and the structure of such flows.An alternate and insightful method entails converting these equations into a set of ordinary first-order autonomous dynamical systems. The analysis of accretion dynamics has shown this formulation, which is widely used in fluid mechanics, to be particularly useful. \citep{jordan1999nonlinear,strogatz1994nonlinear,1993JFM...254..635B,2005astro.ph.11018R,2003MNRAS.344...83R,2003ApJ...592..354A,2006MNRAS.373..146C,2012NewA...17..285N,2003MNRAS.344...83R,2002PhRvE..66f6303R,2018MNRAS.480.3017M}. In order to apply the above method, a coupled system of first-order differential equations can be constructed by expressing the stationary polytropic flow equations (see equation~\ref{cri1}) in a parametrised form.
\begin{equation}
\begin{aligned}
\frac{\mathrm{d}v^{2}}{\mathrm{d}\tau} &= \mathcal{N},
\qquad
\frac{\mathrm{d}r}{\mathrm{d}\tau} &= \mathcal{D}.
\end{aligned}
\end{equation}
where $\tau$ is an arbitrary mathematical parameter. It is important to note that in both cases, the parameter \( \tau \) does not explicitly appear on the right-hand side. Expanding the dynamical variables about their critical (fixed–point) values,
we introduce small perturbations as
\begin{equation}
\begin{aligned}
v^{2} &= v_{c}^{2} + \delta v^{2}, \\
c_{s}^{2} &= c_{sc}^{2} + \delta c_{s}^{2}, \\
r &= r_{c} + \delta r ,
\end{aligned}
\end{equation}
where $\delta v^{2}$, $\delta c_{s}^{2}$ and $\delta r$ represent small
perturbations about the critical point $(v_c^2, c_{sc}^2, r_c)$.

one could derive a set of two autonomous first-order linear differential equations
in the $\delta r$-$\delta v^2$  plane, with $\delta c_{\mathrm{s}}^{2} $
 having to be first expressed in terms of $\delta r$ and $\delta v^2$.
\begin{equation}
\delta c_{\mathrm{s}}^{2} = -c_{\mathrm{sc}}^{2} (\gamma - 1) \left\{
\frac{\delta v^{2}}{2 v_{\mathrm{c}}^{2}} +
\left[1 + r_{\mathrm{c}} \frac{\overset{\boldsymbol{\cdot}}{H}\left(r_{\mathrm{c}}\right)}{H\left(r_{\mathrm{c}}\right)}\right]
\frac{\delta r}{r_{\mathrm{c}}}
\right\}
\end{equation}
The resulting coupled set of linear equations in $\delta r$ and $\delta v^2$ will follow simply as
\begin{equation}
\begin{aligned}
\frac{\mathrm{d}}{\mathrm{d}\tau}(\delta v^{2}) &=
\mathcal{P}\,\delta v^{2} + \mathcal{Q}\,\delta r, \\
\frac{\mathrm{d}}{\mathrm{d}\tau}(\delta r) &=
\mathcal{R}\,\delta v^{2} + \mathcal{S}\,\delta r .
\end{aligned}
\end{equation}

\begin{equation}
\begin{aligned}
\mathcal{P} &= -(\gamma - 1)\left[ 1 + r_{\mathrm{c}}
\frac{\dot H(r_{\mathrm{c}})}{H(r_{\mathrm{c}})} \right] c_{\mathrm{sc}}^{2}, \\[6pt]
\mathcal{Q} &= - 2 c_{\mathrm{sc}}^{2} \Bigg[
\frac{2 \lambda^{2}}{r_{\mathrm{c}}^{3}}
+ \dot{\Phi}^{\mathrm{Gal}}_{i}(r_{\mathrm{c}})
+ r_{\mathrm{c}} \ddot{\Phi}^{\mathrm{Gal}}_{i}(r_{\mathrm{c}})  \\
&\qquad + (\gamma - 1)
\left\{ 1 + r_{\mathrm{c}} \frac{\dot H(r_{\mathrm{c}})}{H(r_{\mathrm{c}})} \right\}^{2}
\frac{c_{\mathrm{sc}}^{2}}{r_{\mathrm{c}}}
- c_{\mathrm{sc}}^{2}
\left\{ (\ln H(r_{\mathrm{c}}))' +
r_{\mathrm{c}} (\ln H(r_{\mathrm{c}}))'' \right\}
\Bigg], \\[6pt]
\mathcal{R} &= \left(\frac{\gamma+1}{2}\right) r_{\mathrm{c}}, \\[6pt]
\mathcal{S} &= (\gamma-1)\left[ 1 + r_{\mathrm{c}}
\frac{\dot H(r_{\mathrm{c}})}{H(r_{\mathrm{c}})} \right].
\end{aligned}
\end{equation}

The above system can be written in matrix form as
\begin{equation}
\frac{d}{d\tau}
\begin{pmatrix}
\delta v^{2} \\
\delta r
\end{pmatrix}
=
\begin{pmatrix}
\mathcal{P} & \mathcal{Q} \\
\mathcal{R} & \mathcal{S}
\end{pmatrix}
\begin{pmatrix}
\delta v^{2} \\
\delta r
\end{pmatrix}.
\end{equation}

Assuming perturbative solutions of the form
\(\delta v^2 \sim \exp(\Omega \tau)\) and
\(\delta r \sim \exp(\Omega \tau)\) \citep{2012NewA...17..285N},
the eigenvalues \(\Omega\) governing the growth of the perturbations
are obtained from

\begin{equation}
\left|
\begin{matrix}
\mathcal{P}-\Omega & \mathcal{Q} \\
\mathcal{R} & \mathcal{S}-\Omega
\end{matrix}
\right| = 0 .
\end{equation}

Expanding the determinant gives

\begin{equation}
\Omega^{2} - (\mathcal{P}+\mathcal{S})\Omega
+ (\mathcal{P}\mathcal{S}-\mathcal{Q}\mathcal{R}) = 0 .
\end{equation}

For the present dynamical system, the trace of the stability matrix
vanishes at the critical point, i.e.
\begin{equation}
\mathcal{P}+\mathcal{S}=0 ,
\end{equation}
and hence the eigenvalue equation reduces to

\begin{equation}
\Omega^{2} = \mathcal{Q}\mathcal{R} - \mathcal{P}\mathcal{S}.
\end{equation}

%
%
The form of $\Omega^2$ can also be used to predict the type of potential critical points.  The critical point will be a saddle point if $\Omega^2 > 0$.  However, it will be a center-type point if $\Omega^2 < 0$, with $\Omega^2$ having the real values in both situations.  Using \citep{2012NewA...17..285N}, we can find the eigenvalues $\Omega^2$ for isothermal flows simply by setting $\gamma = 1$.
Investigating possibilities for shock transitions inside the flow is especially important after identifying the stationary flow solutions and the corresponding sonic points (see \S~\ref{flowD}).  The distribution of sonic points shows whether the flow can travel across across transonic regimes smoothly, whereas the stationary solutions outline the accretion system's global structure.  When there are several sonic points, the flow may undergo a discontinuous but physically acceptable transition between two different transonic branches, known as a shock.  The aforementioned phenomenon is investigated in detail in the following section.
\subsection{Shock Formation in Axisymmetric Inviscid Accretion}
\label{shok-F}
In the present work, we analyse shock transitions supported by angular momentum in axisymmetric, inviscid accretion flows under galactic potential.  Rankine-Hugoniot types are required for adiabatic flow that conserves energy across shock.  The Rankine-Hugoniot conditions can then be expressed as \citep{1959flme.book.....L},
\begin{equation}
\begin{split}
[[\rho v]] &= 0 \\
[[p + \rho v^2]] &= 0 \\
\left[\left[ \dfrac{v^2}{2} + h \right]\right] &= 0 \label{RH3}
\end{split}
\end{equation}
These represented for the conservation of momentum, energy, and mass both before and after the shock.  The discontinuity in any flow variable across the shock surface is indicated by the bracketed terms in equation \eqref{RH3}., i.e., \[[[\text{Shock variables}]] = \text{Shock variables}_1 - \text{Shock variables}_2,\] , with $1$ and $2$ representing values on the two sides of the shock front.  for flows in vertical equilibrium geometry, the corresponding shock condition becomes \citep{2016NewA...43...10S},
\begin{equation}
\left[\left[ \dfrac{[M^2(3\gamma - 1) + 2]^2}{2M^2 + (\gamma - 1)M^4} \right]\right] = 0.
\end{equation}
These formulas serve as the fundamental criterion for shock detection by relating the upstream and downstream Mach numbers across the shock.

 Since energy is not conserved in isothermal shocks, they are dissipative in nature.  Additionally, the shock conditions in this case can be written as
\begin{equation}
\begin{split}
\rho_+ v_+ &= \rho_- v_- \\
P_+ + \rho_+ v_+^2 &= P_- + \rho_- v_-^2 \\
M_+ + \dfrac{1}{M_+} &= M_- + \dfrac{1}{M_-} \\ \label{disshok}
\end{split}
\end{equation}
We use the subscripts `$-$' and `$+$' to denote pre-shock and post-shock quantities, respectively. In the upcoming section, we discuss the stability of the stationary solutions obtained.
Investigating the stationary solutions' stability under small perturbations becomes essential once they have been obtained (\S~\ref{flowD}).  The flow behaviour can be greatly affected by very small disturbances, which could cause the steady-state configuration to become unstable.  We apply time-dependent perturbations to the continuity and Euler equations in order to determine whether the transonic solutions and shocks still remain dynamically viable.
In the following section, we present a detailed stability analysis of these stationary configurations.

\subsection{Time-Dependent Stability Analysis of Stationary Solutions}
\label{stable-F}

We consider a galactic potential and obtain the flow behaviour in an AGN considering stationary mode. To ensure the stability of these stationary solutions, one must check the validity by considering the time dependent flow dynamics \citep{2009MNRAS.398..841B,2006MNRAS.373..146C,2007MNRAS.378.1407G,2012NewA...17..285N,2018MNRAS.480.3017M,1980MNRAS.191..571P,1992ApJ...384..587T}. In this regard the governing equations i.e. the time dependent Euler equation and continuity equation can be written as,

\begin{equation}
\frac{\partial v}{\partial t} + v \frac{\mathrm{d} v}{\mathrm{d} r} + \frac{1}{\rho} \frac{\mathrm{d} P}{\mathrm{d} r} + \overset{\boldsymbol{\cdot}}{\Phi}{}^{\text{Gal}}_{i}(r) - \frac{\lambda^{2}}{r^{3}} = 0
\label{euler11}
\end{equation}

\begin{equation}
\frac{\partial \rho}{\partial t} + \frac{1}{r H} \frac{\partial}{\partial r}
\left( \rho v r H \right) = 0 \,.
\label{cont111}
\end{equation}
We introduced here a time dependent perturbative analysis to check the time viability of the solutions and expressed the perturbed quantities as, $v(r,t) = v_0(r) + \tilde{v}(r,t)$ and
$\rho (r,t) = \rho_0 (r) + \tilde{\rho}(r,t)$. From equ \eqref{cont111}, we introduce the variable \( f = \rho v r H \) whose stationary value is a constant \( f_0 \) also known as matter flux rate, yields us a relation that connects all the three
fluctuating quantities, $\tilde{v}$,  $\tilde{\rho}$
and $\tilde{f}$, with one another and given by,
\begin{equation}
\label{effprimeh}
\frac{\tilde{f}}{f_0} = \frac{\tilde{\rho}}{\rho_0}
+ \frac{\tilde{v}}{v_0}
\end{equation}
where $\sim$ symbol denotes the small time-dependent perturbations and quantities with subscript 0 denotes stationary background flow variables. For VE model, since the disc height $H$ depends on both the density $\rho$ and the radial coordinate $r$, the above equation \eqref{cont111} get modified to \citep{2006MNRAS.373..146C,2012NewA...17..285N},
\begin{equation}
\label{volden}
\frac{\partial}{\partial t} \left[\rho^{(\gamma +1)/2}\right]
+ \frac{\sqrt{\overset{\boldsymbol{\cdot}}{\Phi}{}^{\text{Gal}}_{i}(r)}}{r^{3/2}}
\frac{\partial}{\partial r} \left[ \rho^{(\gamma +1)/2} \, v \,
\frac{r^{3/2}}{\sqrt{\overset{\boldsymbol{\cdot}}{\Phi}{}^{\text{Gal}}_{i}(r)}} \right] = 0 \,,
\end{equation}
where
\begin{equation}
f=\frac{\rho^{(\gamma +1)/2} \, v \, r^{3/2}}{\sqrt{\overset{\boldsymbol{\cdot}}{\Phi}{}^{\text{Gal}}_{i}(r)}} \,,
\end{equation}
and we can write,
\begin{equation}
\label{effprime}
\frac{\tilde{f}}{f_0} = \left( \frac{\gamma +1}{2} \right)
\frac{\tilde{\rho}}{\rho_0} + \frac{\tilde{v}}{v_0} \,.
\end{equation}
Therefore, equ \eqref{volden}, now beocmes
\begin{equation}
\label{flucden}
\frac{\partial \tilde{\rho}}{\partial t} + \beta^2
\frac{v_0 \rho_0}{f_0} \left(\frac{\partial \tilde{f}}{\partial r}\right)=0 \,,
\end{equation}
where $\beta^2 = \frac{2}{\gamma +1}$. The velocity fluctuations can be written as
\begin{equation}
\label{flucvel}
\frac{\partial \tilde{v}}{\partial t}= \frac{v_0}{f_0}
\left(\frac{\partial \tilde{f}}{\partial t}+ v_0
\frac{\partial \tilde{f}}{\partial r}\right) \,
\end{equation}
which, upon a further partial differentiation with respect to time,
will give
\begin{equation}
\label{flucvelder2}
\frac{{\partial}^2 \tilde{v}}{\partial t^2}=\frac{\partial}{\partial t} \left[
\frac{v_0}{f_0} \left(\frac{\partial \tilde{f}}{\partial t}\right) \right]
+ \frac{\partial}{\partial t} \left[ \frac{v_0^2}{f_0} \left(
\frac{\partial \tilde{f}}{\partial r}\right) \right] \,
\end{equation}
The linearised fluctuating part from the time-dependent Euler equation (\ref{euler11})
can be written as,
\begin{equation}
\label{fluceuler}
\frac{\partial \tilde{v}}{\partial t}+ \frac{\partial}{\partial r}
\left( v_0 \tilde{v} + c_{\mathrm{s0}}^2
\frac{\tilde{\rho}}{\rho_0}\right) =0 \,
\end{equation}
with $c_{\mathrm{s0}}$ being the speed of sound in the steady state. Differentiating Eq.~(\ref{fluceuler}) with respect to $t$, and Eq.\eqref{flucden}) with the velocity fluctuations and its derivative  we get,
\begin{multline}
\label{interm}
\frac{\partial}{\partial t} \left[\frac{v_0}{f_0}
\left( \frac{\partial \tilde{f}}{\partial t} \right) \right]
+ \frac{\partial}{\partial t} \left[\frac{v_0^2}{f_0}
\left( \frac{\partial \tilde{f}}{\partial r} \right) \right] \\
+ \frac{\partial}{\partial r} \left[\frac{v_0^2}{f_0}
\left( \frac{\partial \tilde{f}}{\partial t} \right) \right]
+ \frac{\partial}{\partial r} \left[\frac{v_0}{f_0}
\left(v_0^2 - \zeta c_{\mathrm{s0}}^2 \right)
\frac{\partial \tilde{f}}{\partial r} \right] = 0
\end{multline}
where $\zeta = \beta^2$ for VE model. The Eq.~(\ref{interm}) can be written in a compact format as,
\begin{equation}
\label{compact}
\partial_\mu \left( {\mathrm{f}}^{\mu \nu} \partial_\nu
\tilde{f}\right) = 0 \,
\end{equation}
where $\mu,\nu =t,r$,
We can write the above equation using a symmetric matrix as,
\begin{equation}
\label{matrix}
{\mathrm{f}}^{\mu \nu } = \frac{v_0}{f_0}
\begin{pmatrix}
1 & v_0 \\
v_0 & v_0^2 - \zeta c_{\mathrm{s0}}^2
\end{pmatrix} \,.
\end{equation}

In curved spacetime, the d'Alembertian (wave operator) acting on a scalar field \( \phi \) is given by \citep{1998CQGra..15.1767V}
\[
\square \phi = \nabla^\mu \nabla_\mu \phi = \frac{1}{\sqrt{-g}} \, \partial_\mu \left( \sqrt{-g} \, g^{\mu\nu} \, \partial_\nu \phi \right),
\]
where \( g^{\mu\nu} \) is the inverse of the metric tensor \( g_{\mu\nu} \), and \( g = \det(g_{\mu\nu}) \). By defining a tensor density \( f^{\mu\nu} = \sqrt{-g} \, g^{\mu\nu} \), this operator simplifies to
\[
\square \phi = \frac{1}{\sqrt{-g}} \, \partial_\mu \left( f^{\mu\nu} \, \partial_\nu \phi \right),
\]

So the metric determinant can be expressed as \( g = \det(f^{\mu\nu}) \), directly linking the density form of the wave operator to the spacetime geometry. It is immediately possible to set down an effective metric for the propagation of an acoustic disturbance as
\begin{equation}
\label{metric}
\mathrm{g}^{\mu \nu}_{\mathrm{eff}} =
\begin{pmatrix}
1 & v_0 \\
v_0 & v_0^2 - \zeta c_{\mathrm{s0}}^2
\end{pmatrix} \,,
\end{equation}
This give $v_0^2 = \zeta c_{\mathrm{s0}}^2$ as the horizon condition of an acoustic black hole for inflow solutions. Finally, a readjustment of terms in Eq.~(\ref{interm}) will
give the following perturbation equation \citep{2006MNRAS.373..146C,2007MNRAS.378.1407G,2012NewA...17..285N,2018MNRAS.480.3017M}:
\begin{equation}
\label{eq:wave_eq}
\frac{\partial^2 \tilde{f}}{\partial t^2}
+ 2 \frac{\partial}{\partial r} \left( v_0 \frac{\partial \tilde{f}}{\partial t} \right)
+ \frac{1}{v_0} \frac{\partial}{\partial r} \left[ v_0 \left(v_0^2 - \zeta c_{\mathrm{s0}}^2 \right)
\frac{\partial \tilde{f}}{\partial r} \right] = 0 \,,
\end{equation}
An analogue event horizon forms at the sonic point as a result of the propagation of acoustic perturbations in a classical, inviscid, and spatially variable transonic fluid.  These points appear together to form a sonic surface, which acts as a border that traps outgoing phonons.  Similar to the causal structure of a black hole event horizon, the flow becomes supersonic beyond this surface, preventing acoustic disturbances carried by the fluid from propagating upstream.  The strength of the acoustic horizon is determined by the velocity gradient at this sonic surface, which functions as an equivalent of the black hole's surface gravity.Acoustic geometry in supersonic fluids was initially introduced by Unruh \citep{1981PhRvL..46.1351U}, who also showed that an analogue surface gravity could be derived for such horizons.  The acoustic horizon is expected to produce Hawking-like thermal radiation in the form of phonons, matching the Hawking radiation from astronomical black holes, which is a striking result of this analogy.
We study the stability characteristics of the acoustic surface gravity in the next section.
\subsection{ACOUSTIC SURFACE GRAVITY}
\label{Surgrav}
The acoustic surface gravity $\kappa$ for the stationary background fluid accreting under the influence of a post-Newtonian black hole potential can be obtained as \citep{1998CQGra..15.1767V,1999PThPS.136....1J,1981PhRvL..46.1351U,2014CQGra..31c5002B,2016NewA...43...10S}
\begin{multline}
\kappa = \left| \sqrt{(1 + 2\Phi^{\text{Gal}}_i(r))
\left( 1 - \frac{\lambda^{2}}{r^{2}}
- 2\Phi^{\text{Gal}}_i(r) \frac{\lambda^{2}}{r^{2}} \right) } \right. \\
\left. \times \left( \frac{1}{1 - c^{2}_{sc}}
\left[ \left. \frac{dv}{dr} \right|_{r_{c}}
- \left. \frac{dc_{s}}{dr} \right|_{r_{c}} \right] \right) \right|
\label{eq.501}
\end{multline}
The surface gravity $\kappa \equiv \kappa[\gamma, \mathcal{E}, \lambda]$. The bracketted terms indicates its functional dependence. We aim to investigate how the acoustic surface gravity depends galactic parameter $\Upsilon_{B} $. For this purpose, we calculate $\kappa$ of $\Phi^{\text{Gal}}_i$ for different values of $\Upsilon_{B} $ while varying galactic parameter, keeping the other parameters $[\gamma, \mathcal{E}, \lambda]$ fixed.

The transonic nature and dynamical stability of accretion flows under the different pseudo-Newtonian potentials, $\Phi^{\mathrm{BH}}_i$, and the galactic potential,
$\Phi^{\mathrm{Gal}}_i$, are investigated in this study.  We analyse the behaviour of the flow under two thermodynamic frameworks: isothermal and adiabatic.  A number of interesting phenomena are revealed by evaluating the prerequisites for shock production in both regimes.  We also compute the equivalent surface gravity connected to the emerging acoustic structure.  Notably, compared to models that simply take into account a central black hole, the addition of galactic-scale gravitational components affects modifications to the locations of critical points and shock fronts.  Despite their small magnitude, these changes have significant physical significance.  The following plots provide detailed illustrations and discussions of all the results and their consequences.

\section{Results \& Discussions}
\label{Result}

In this section, we present a comparative analysis of axisymmetric, low–angular-momentum accretion flows in the presence of a multi-component galactic potential, where the central black hole (BH) gravity is modelled using four different pseudo-Schwarzschild potentials. The vertical equilibrium (VE) disc model is adopted throughout, and the accreting fluid is described using two distinct thermodynamic equations of state. To examine how the ambient galactic environment influences the accretion dynamics, we incorporate the combined gravitational contributions from three major astrophysical components: (i) the surrounding stellar distribution, (ii) the dark matter halo, and (iii) the diffuse hot ionized interstellar gas, in addition to the pseudo-Schwarzschild potential representing the central BH \citep{2005MNRAS.363..705M,2018MNRAS.479.3011R,2021JCAP...05..025R,2025arXiv250918833S}.

Our primary objective is to investigate how the locations and properties of the critical points, the shock formation criteria, and the resulting transonic features vary under these different configurations. As discussed earlier, the presence of a standing shock introduces a discontinuous compression in the flow, thereby modifying the mass accretion rate by altering the post-shock density and velocity profiles. Previous studies have shown that inward shock propagation strengthens the inflow through enhanced post-shock compression, while outward propagation weakens the post-shock confinement and may initiate outflowing motion. Such shock-driven perturbations are capable of inducing measurable fluctuations in the effective accretion rate near the event horizon \citep{1996ApJ...457..805M,2001ApJ...546..429B,2009ApJ...702..649D}.

Motivated by these considerations, we aim to identify how the shock and sonic locations change when different prescribed BH potentials are embedded within the broader galactic potential. We explore the full parameter space to quantify these differences and also analyse the nature of the critical points using the eigenvalue stability matrix. The shock induced variables were also investigated and  the corresponding effective acoustic surface gravity for each case is likewise computed.

In the following subsections, we present our results and discuss each figure in detail, beginning with the characteristics of the critical points.

\subsection{Critical point analysis and multi-critical parameter space}
\begin{figure*}
  \centering
  \begin{subfigure}[b]{0.48\linewidth}
    \includegraphics[width=\linewidth,clip]{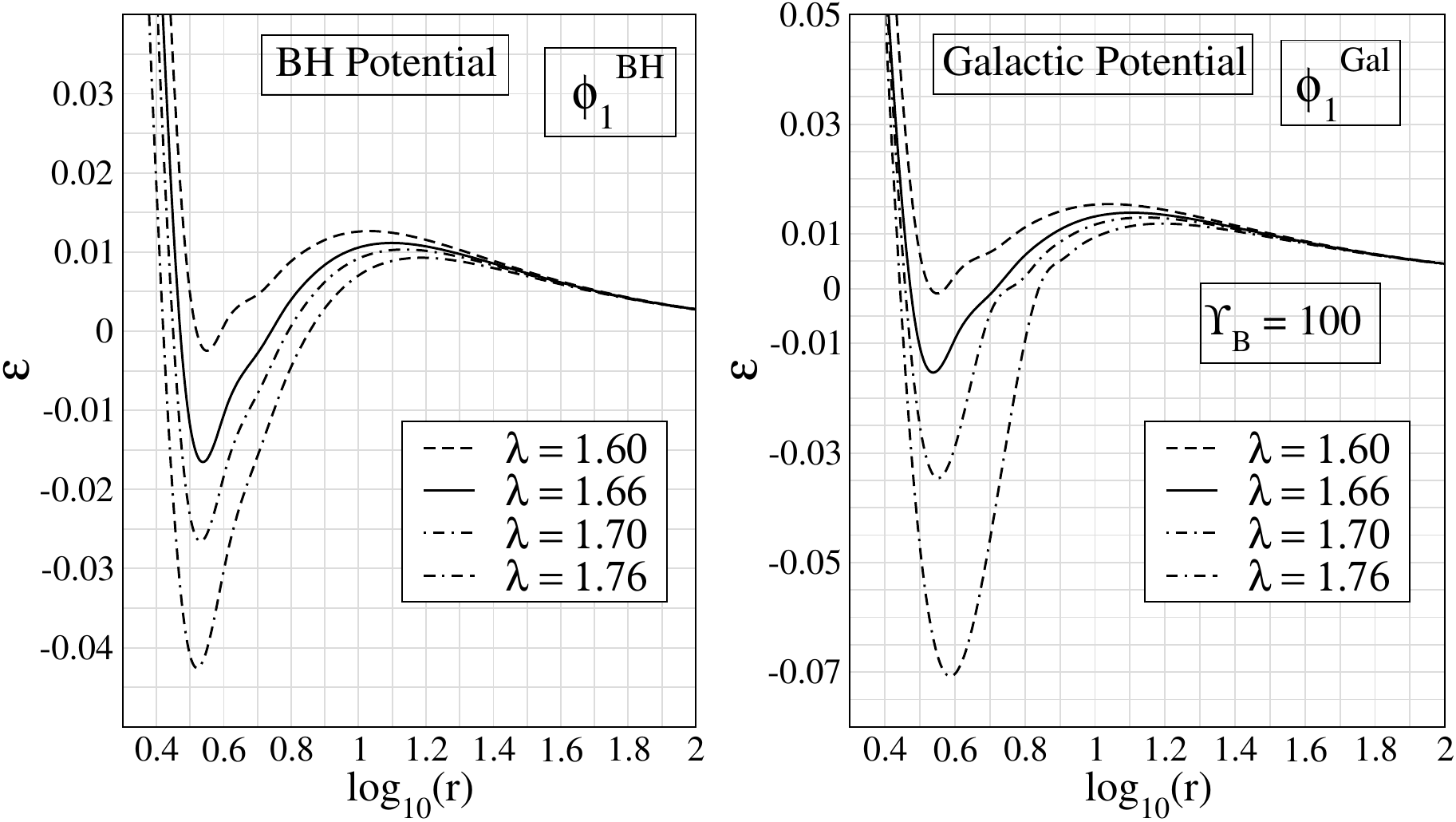}
    \caption{}
    \label{phi1a}
  \end{subfigure}
  \begin{subfigure}[b]{0.48\linewidth}
    \includegraphics[width=\linewidth,clip]{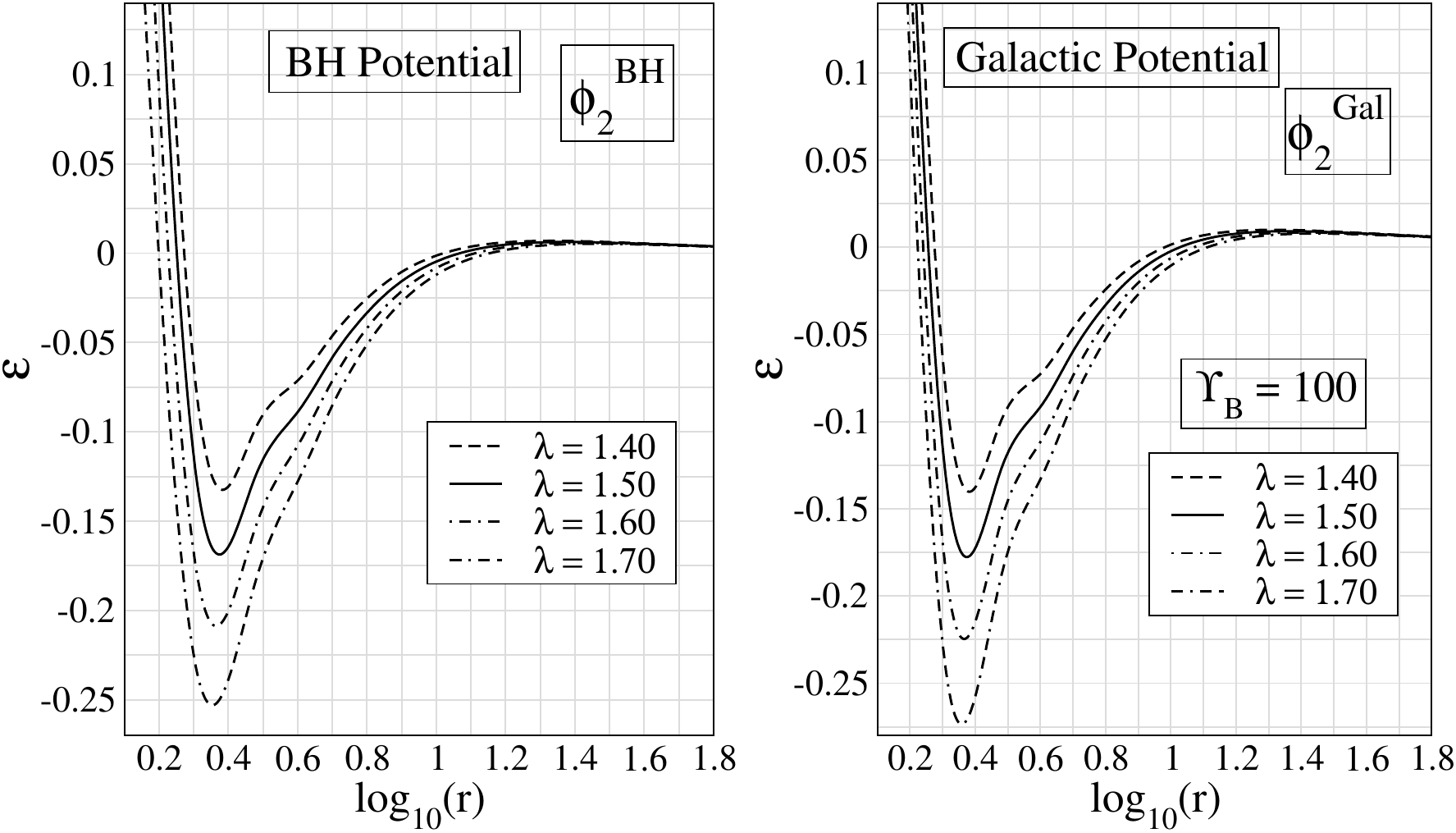}
    \caption{}
    \label{phi2b}
  \end{subfigure}

  \begin{subfigure}[b]{0.48\linewidth}
    \includegraphics[width=\linewidth,clip]{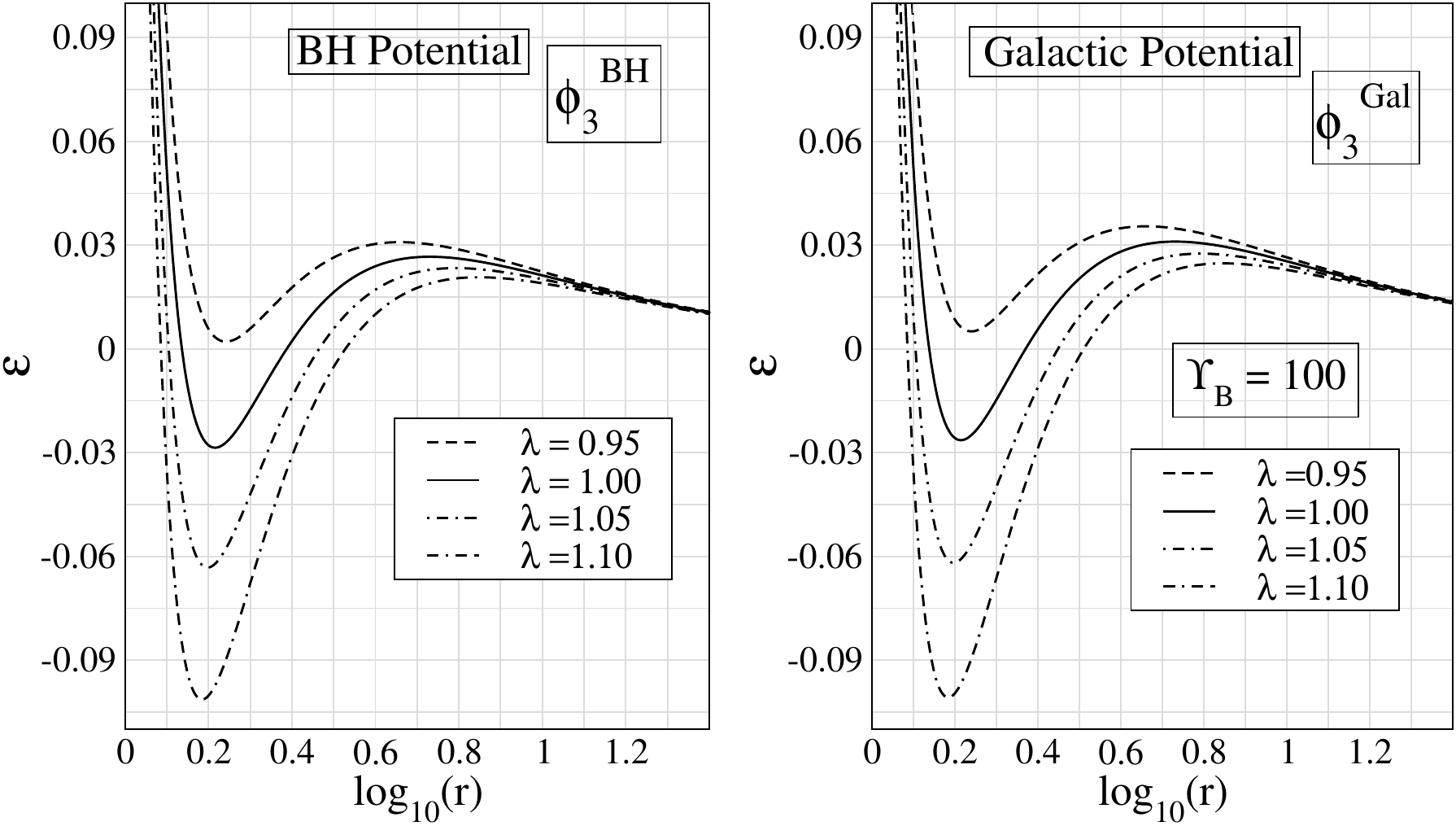}
    \caption{}
    \label{phi3c}
  \end{subfigure}
  \begin{subfigure}[b]{0.48\linewidth}
    \includegraphics[width=\linewidth,clip]{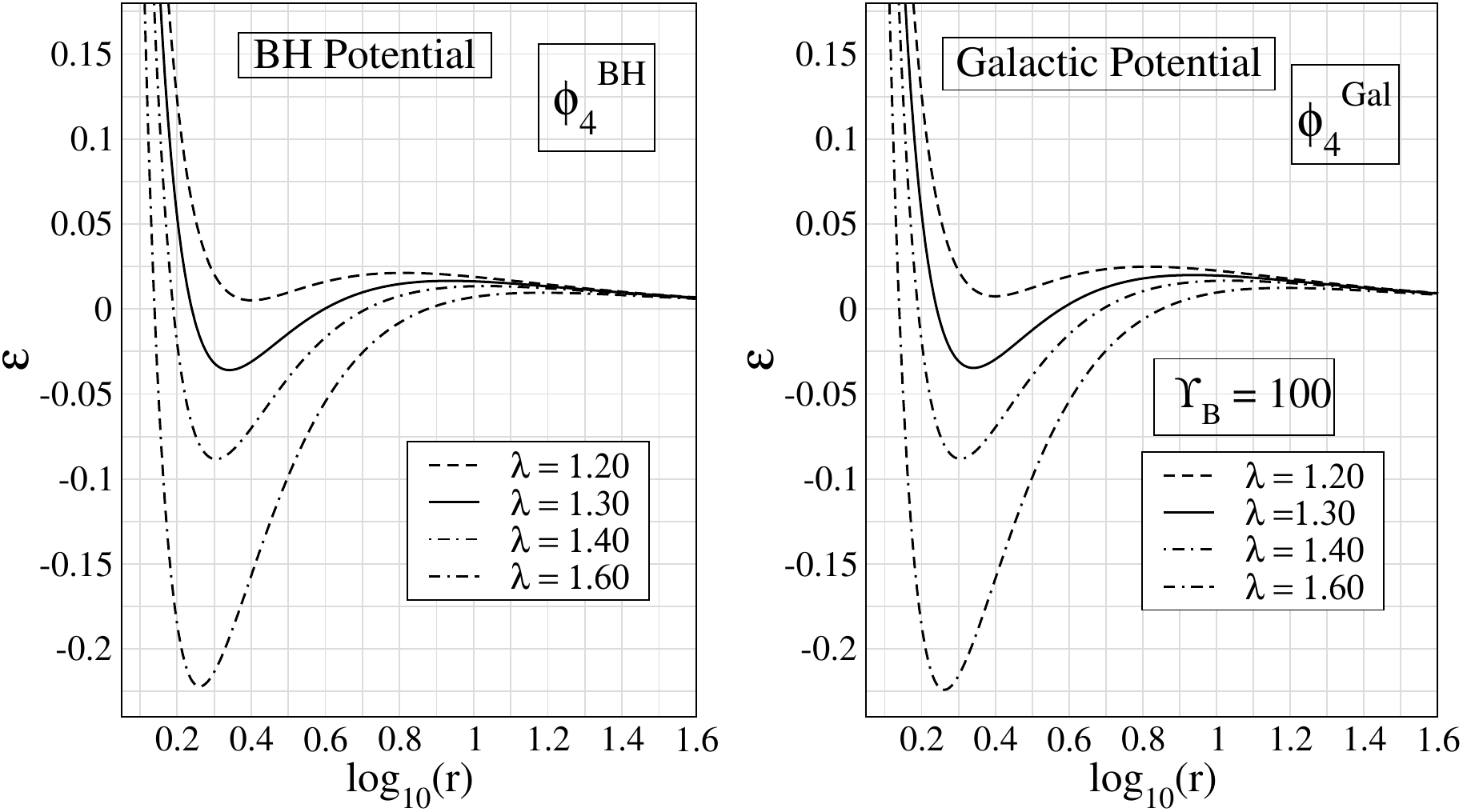}
    \caption{}
    \label{phi4d}
  \end{subfigure}
\caption{In this figure we show the variation of the specific energy of the flow with the location of critical points for different values of specific angular momentum, using $\Phi^{\text{BH}}_i$ for the black hole pseudo potential and $\Phi^{\text{Gal}}_i$ for the galactic potential. The curves are drawn for different values of $\lambda$: (a) $\lambda = 1.60,\, 1.66,\, 1.70,\, 1.76$; (b) $\lambda = 1.40,\, 1.50,\, 1.60,\, 1.70$; (c) $\lambda = 0.95,\, 1.00,\, 1.05,\, 1.10$; and (d) $\lambda = 1.20,\, 1.30,\, 1.40,\, 1.60$. The critical points are obtained as the intersections of the constant-energy (horizontal) lines with the constant-angular-momentum curves. The number of intersections of these horizontal lines with the plotted curves indicates the number of critical points for the corresponding flow configurations. The details are discussed in text and positions of inner ,middle and outer points along with their respective energy and angular momentum are given in Table \ref{tab:critical-points-VE}. }
\label{E-r}
\end{figure*}
\begin{figure*}
  \centering
  \begin{subfigure}[b]{0.45\linewidth}
    \includegraphics[width=\linewidth,clip]{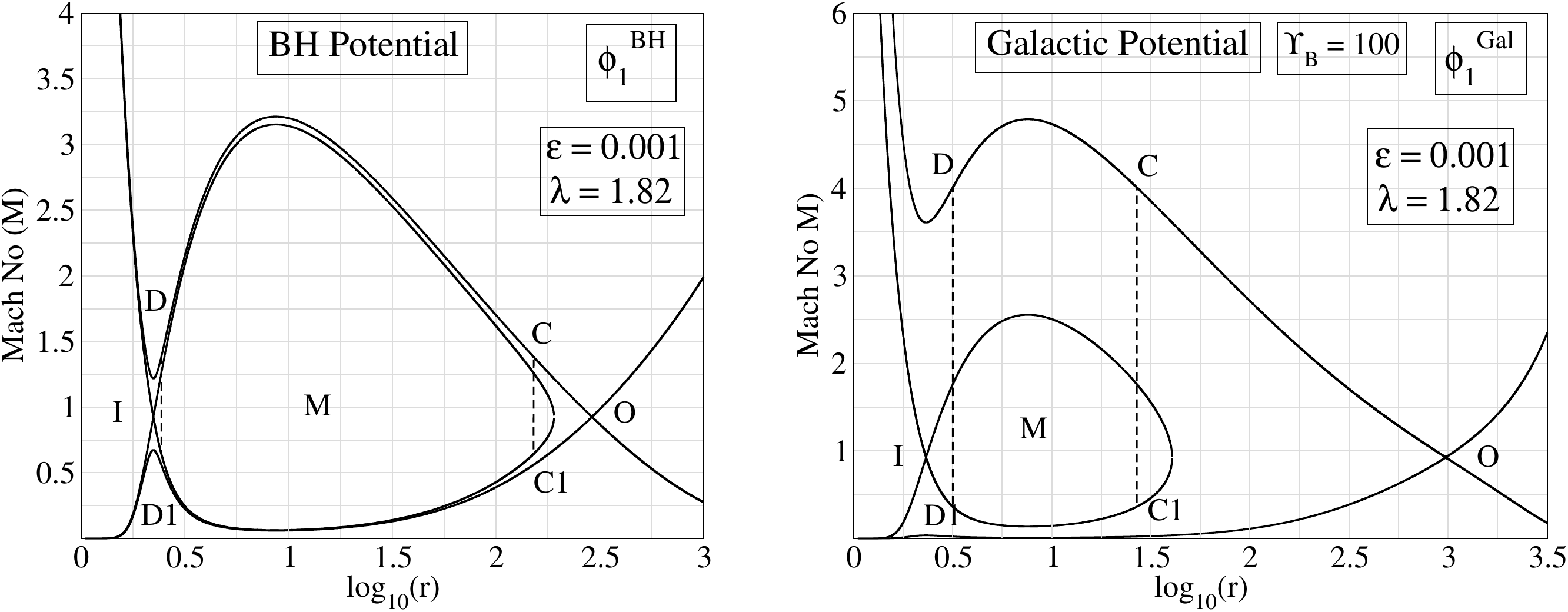}
    \caption{}
    \label{VE1a}
  \end{subfigure}
  \begin{subfigure}[b]{0.45\linewidth}
    \includegraphics[width=\linewidth,clip]{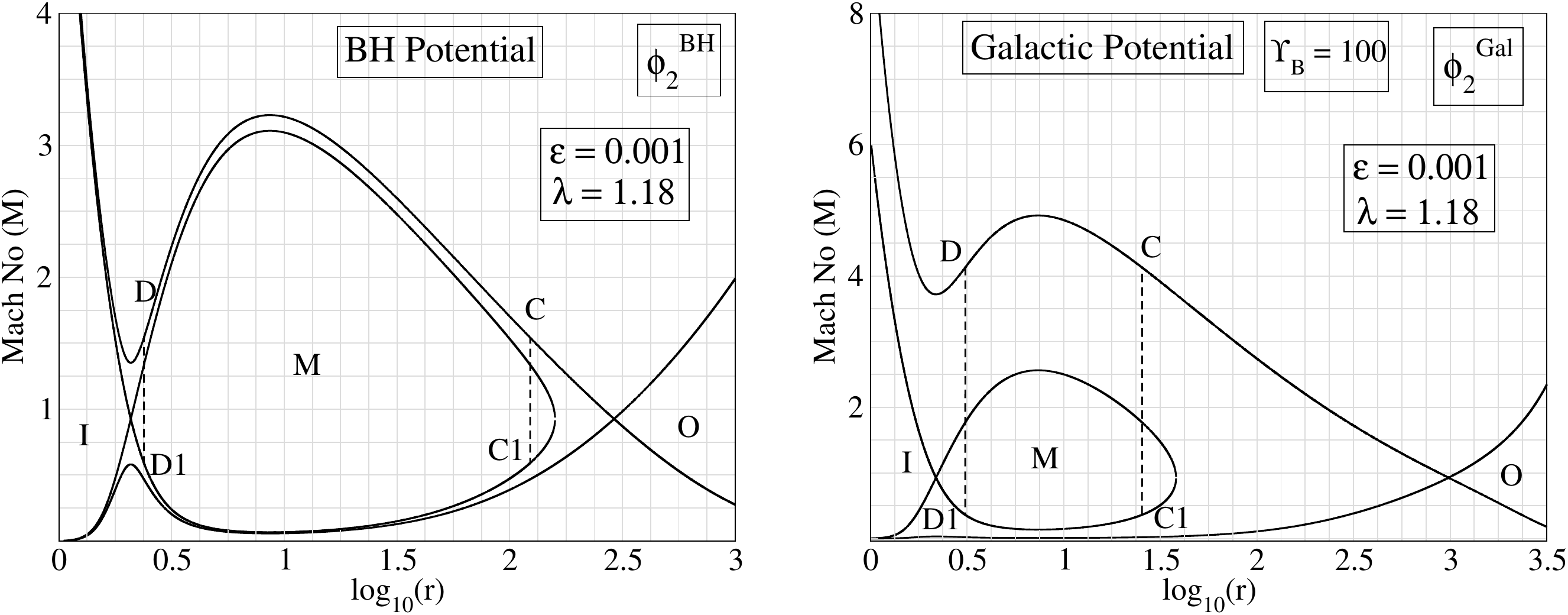}
    \caption{}
    \label{VE2b}
  \end{subfigure}

  \begin{subfigure}[b]{0.45\linewidth}
    \includegraphics[width=\linewidth,clip]{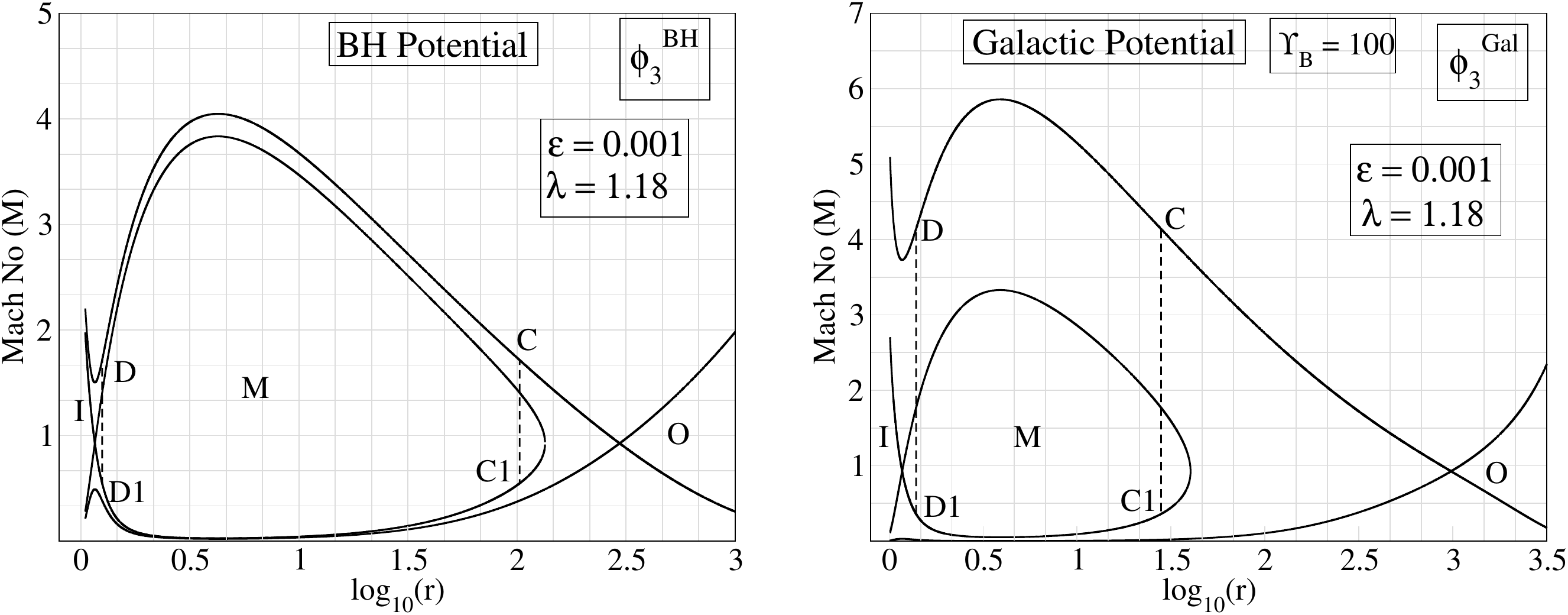}
    \caption{}
    \label{VE3c}
  \end{subfigure}
  \begin{subfigure}[b]{0.45\linewidth}
    \includegraphics[width=\linewidth,clip]{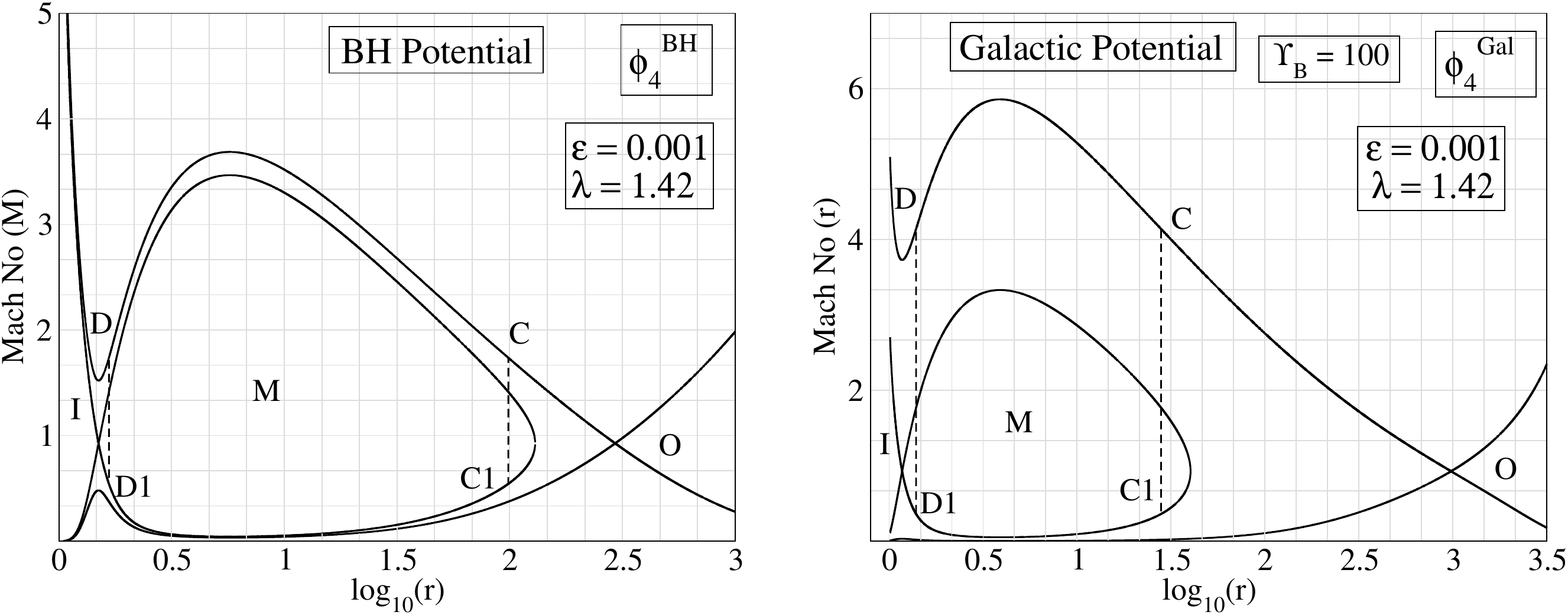}
    \caption{}
    \label{VE4d}
  \end{subfigure}
   \caption{ We present here the phase portrait of adiabatic transonic accretion where the horizontal axis represent the $log_{10}r$ and the vertical axis represent the mach number $M=\frac{v}{c_s}$. $r$ is the radial distance from the central black hole, $v$ and $c_s$ are the flow veocity and sound speed respectively. The left panels show the plots using BH pseudo potential $\Phi^{\mathrm{BH}}_i$, while right panels include  galactic potential $\Phi^{\mathrm{Gal}}_i$ with galactic  parameter $\Upsilon_B=100$. Inner ($I$), middle ($M$) and outer ($O$) sonic points are marked. $CC1$ and $DD1$ denote two measures of the outer and inner discontinuous shock transition in the flow. Different parameter sets and computed quantities for the four cases (panels a--d) are discussed in Table \ref{phase-tab}. The respective energy and angular momentum are given in associated legend box. In all four cases, $\Phi^{\mathrm{Gal}}_i$ pushes the outer sonic radius $O$ outward, demonstrating that the large-scale behaviour is dominated by the extended galactic gravitational field rather than by the black--hole potential alone. We also see that the outer shock location greatly influenced in the presence of $\Phi^{\mathrm{Gal}}_i$, it shift inward indicating the compression of post shock region in all the four cases.
   Moreover, the shock-strength measure $CC1$ is noticebly increased in the presence of $\Phi^{\mathrm{Gal}}_i$,  while $DD1$ typically remain same. These highlight that the galactic potential reshapes the outer transonic topology and alters the character of the discontinuity, even while leaving the inner critical region approximately intact. The inclusion of the galactic potential has an important impact on the global transonic structure of the flow. }
  \label{PP1}
\end{figure*}

First, using four distinct pseudo-Schwarzschild potentials in the figure \ref{E-r}, we explore the transonic feature of the accretion flow under the galactic potential and compare it with the BH potential. Using $\Phi^{\text{BH}}_i$ for the black hole potential and $\Phi^{\text{Gal}}_i$ for the galactic potential, taking into account $\Upsilon_B = 100$, these illustrate how the specific energy ( $\mathcal{E}$) of the flow varies with the location of the critical points for varying values of the specific angular momentum ($\lambda$). The curves are drawn for different values of $\lambda$: (a) $\lambda =1.60,\,1.66,\,1.70,\,1.76$; (b) $\lambda =1.40,\,1.50,\, 1.60,\,1.70$; (c) $\lambda = 0.95,\,1.00,\,1.05,\,1.10$; and (d) $\lambda = 1.20,\,1.30,\,1.40,\,1.60$. The constant-energy (horizontal) lines' intersections with the constant-angular-momentum curves yield the critical points.  The number of critical points associated with a specific flow configuration can be calculated by the number of intersections between these horizontal lines and the corresponding curves.
 Three critical points can be identified in each case.  The inner and outer sonic points are classified as \emph{saddle-type} or \emph{X-type} critical points, which correspond to areas where the slope of the $\mathcal{E}$--$r$ curve is negative.  The flow can smoothly transition from subsonic to supersonic speeds (or vice versa) through these physically relevant locations.  The intermediate critical point that prevents such a physical transition is referred to as a \emph{center-type} or \emph{O-type} point. It is located between the two X-type points and is connected to a positive slope.
It can be seen from the results above that the location of the critical points is considerably impacted by the galactic potential $\Phi^{\text{Gal}}_i$. For each potential pair $(\Phi^{\text{BH}}_i,\, \Phi^{\text{Gal}}_i)$, the innermost critical point $r_{c1}$ remains nearly unchanged, since it is primarily governed by the strong-field region of the black hole potential. However, both the middle ($r_{c2}$) and the outer ($r_{c3}$) critical points shift to comparatively larger radii in the presence of the galactic potential. This outward shift arises due to the additional gravitational contribution from the galactic components , which effectively deepens the potential well at larger scales. To further elaborate on the nature of these three types of critical points and to understand their associated phase topologies in more detail, we investigate the phase portrait of a polytropic transonic accretion flow within a VE disc geometry, embedded in a galactic environment, as shown in Figure \ref{PP1}.

 \begin{table*}
  \centering
   \begin{tabular}{@{}|l| l| c| c| c| c| c|@{}}
    \toprule
    \textbf{Panel} & \textbf{Potential} & $\boldsymbol{\mathcal{E}_c}$ & $\boldsymbol{\lambda}$ & $\boldsymbol{r_{c1}(r_g)}$ & $\boldsymbol{r_{c2}(r_g)}$ & $\boldsymbol{r_{c3}(r_g)}$ \\
    \midrule
    (a) & $\Phi^{\mathrm{BH}}_1$  & 0.001 & 1.66 & 2.84 & 5.66 & 292.33 \\
        & $\Phi^{\mathrm{Gal}}_1$ & 0.001 & 1.66 & 2.89 & 5.02 & 677.18 \\
    \midrule
    (b) & $\Phi^{\mathrm{BH}}_2$  & 0.001 & 1.50 & 1.58 & 12.23 & 287.32 \\
        & $\Phi^{\mathrm{Gal}}_2$ & 0.001 & 1.50 & 1.58 & 9.98 & 662.20 \\
    \midrule
    (c) & $\Phi^{\mathrm{BH}}_3$  & 0.001 & 1.00 & 1.36 & 2.47 & 296.95 \\
        & $\Phi^{\mathrm{Gal}}_3$ & 0.001 & 1.00 & 1.38 & 2.27 & 672.70 \\
    \midrule
    (d) & $\Phi^{\mathrm{BH}}_4$  & 0.001 & 1.30 & 1.70 & 4.16 & 294.93 \\
        & $\Phi^{\mathrm{Gal}}_4$ & 0.001 & 1.30 & 1.71 & 3.79 & 674.20 \\
    \bottomrule
  \end{tabular}
  \caption{ Each panel (a--d) corresponds to a distinct parameter set as described in Fig.~\ref{E-r}. The quantities $r_{c1}$, $r_{c2}$, and $r_{c3}$ represent the inner, middle, and outer critical points, respectively, for both  $\Phi^{\mathrm{BH}}_i$ and  $\Phi^{\mathrm{Gal}}_i$ .}
   \label{tab:critical-points-VE}
\end{table*}

\begin{figure*}
 \centerline{\includegraphics[width=0.85\linewidth,clip]{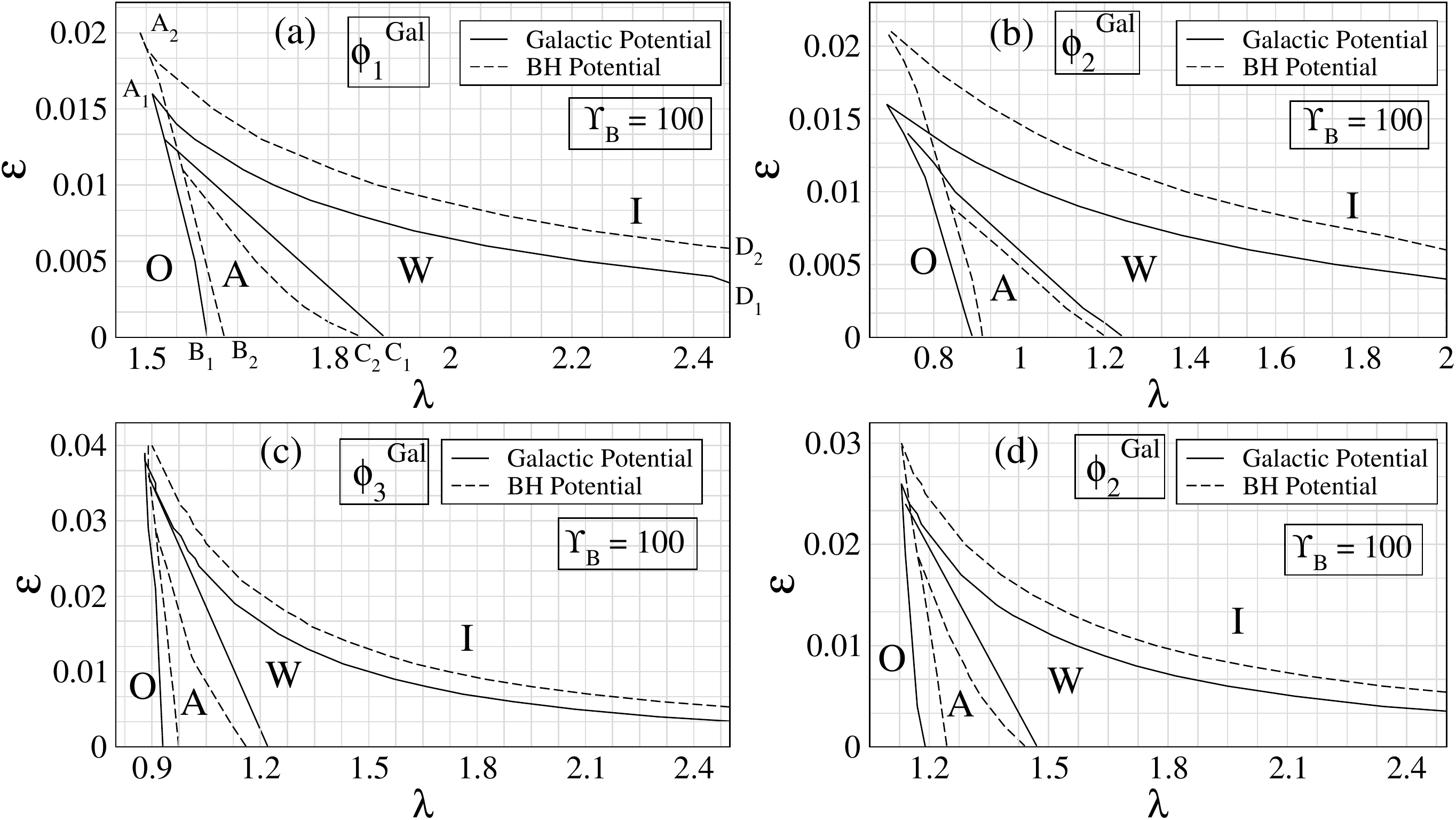}}
  \caption{We plot different regions in the parameter space of specific energy ($\mathcal{E}$) and specific angular momentum ($\lambda$), corresponding to the number and nature of critical points in adiabatic accretion flows. The diagram shows the classification of these regions for the VE disc model. The parameter space is plotted for $\Phi^{\mathrm{BH}}_i$ (dashed line) considering  BH pseudo Schwarzschild potential, and for $\Phi^{\mathrm{Gal}}_i$ representing the  galactic potential (solid line). The galactic parameter is set to $\Upsilon_{B}=100$, which characterizes the degree of inclusion of galactic effects . Each diagram is divided into four distinct regions: $O$, $A$, $W$, and $I$, corresponding to different phase portrait topologies (see text for details). Inclusion of the galactic potential causes the wedge-shaped multitransonic region to shrink and shift towards the centrally located black hole for each potential. The overall structure and topology of the critical point regions vary systematically with  the galactic background, demonstrating how the transonic properties of the flow are  influenced by the galactic environment.
}
\label{PMSCHCOVE}
\end{figure*}

The critical points of the flow are determined by fixing the parameters $f(\mathcal{E},\lambda)$ in Fig.~\ref{E-r}. To determine the whole parameter space (PMS) in the $\mathcal{E}$–$\lambda$ plane where the accretion flow exhibits multi-transonic behavior, we analyze Eq.~\eqref{E-space} and map the admissible region, as seen in Fig.~\ref{PMSCHCOVE}. The dashed line in each subpanel represented $\Phi^{\text{BH}}_i$, whereas the solid line showed $\Phi^{\text{Gal}}_i$. The galactic parameter that we have chosen is $\Upsilon_{B} = 100$.
Depending on the number and nature of the critical points, each panel showing the parameter space is separated into four regions denoted as $\boldsymbol{\textbf{O}}$, $\boldsymbol{\textbf{I}}$, $\boldsymbol{\textbf{A}}$, and $\boldsymbol{\textbf{W}}$.  The flows in the $\boldsymbol{\textbf{O}}$ region have a single outer saddle-type critical point ($r_\text{out}$), which is far from the black hole horizon.  The $\boldsymbol{\textbf{I}}$ area indicates solutions that are closer to the horizon and have a single inner saddle-type critical point ($r_\text{in})$.  On the other hand, flows with three critical points—an inner and an outer saddle-type point ($r_\text{in}$ and $r_\text{out}$), separated by a center-type point ($r_\text{mid}$)—correspond to the $\boldsymbol{\textbf{A}}$ and $\boldsymbol{\textbf{W}}$ regions.  In global accretion solutions, $r_\text{mid}$ has minimal physical relevance because it does not support transonic flow, despite its mathematical existence. The entropy distribution at the critical points that occur when $\boldsymbol{\textbf{A}}$ and $\boldsymbol{\textbf{W}}$ differ. Shock formation in accretion-type flows is possible in the $\boldsymbol{\textbf{A}}$ region because the entropy at the inner critical point is larger than that at the outer one, i.e., $\mathcal{\dot{M}}(r_\text{in}) > \mathcal{\dot{M}}(r_\text{out})$.  On the other hand, in the $\boldsymbol{\textbf{W}}$ region, the entropy favours shock generation in wind-type flows by obeying $\mathcal{\dot{M}}(r_\text{in}) < \mathcal{\dot{M}}(r_\text{out})$.  The middle curve in each panel's wedge-shaped multi-critical region indicates the boundary between these two regions, where entropy values at the two saddle points are equal. For BH potential the multi-critical region is bounded by points $A_2$, $B_2$, and $D_2$, and contains two subregions. The subregion enclosed by $A_2$, $B_2$, and $C_2$ satisfies $\dot{\mathcal{M}}_\text{in} > \dot{\mathcal{M}}_\text{out}$, while the region enclosed by $A_2$, $C_2$, and $D_2$ shows the opposite entropy trend. Along the curve $A_2 C_2$, the entropy at both critical points is identical . Similar wedge-shaped multi-critical zones appear for galactic potential, labeled by $A_1$, $B_1$, and $D_1$ . Thus, for low $\mathcal{E}$, there exists a finite range in $\lambda$ that allows three critical points to form. This makes multi-transonic behavior possible under certain conditions, particularly when $\dot{\mathcal{M}}_\text{in} > \dot{\mathcal{M}}_\text{out}$ and the Rankine–Hugoniot shock conditions are satisfied. Outside the multi-critical wedges, only single critical points are found, limiting the flow to mono-transonic solutions. The transonic nature of accretion processes is profoundly altered by the addition of the galactic potential, as demonstrated by the analysis of the parameter space in the $(\mathcal{E}, \lambda)$ plane. In the pure BH case ($\Phi^{\mathrm{BH}}_i$), the multitransonic region is broader, indicating a wider range of admissible energy–angular momentum combinations that allow multiple critical points. When the  galactic potential ($\Phi^{\mathrm{Gal}}_i$) is included, this region contracts and shifts toward lower $\lambda$ values, signifying that the external galactic gravity stabilizes the flow and reduces the range of conditions under which multiple sonic transitions can occur.

Now, we can easily identify the regions that are part of types I, O, A, and W, and at the same time the $\mathcal{E}$–$\lambda$ domains that correspond right alongside them. However, this analysis does not reveal us anything about the crucial points' nature or classification. Therefore, we choose typical fixed values of $\mathcal{E}$ and $\lambda$  in order to quantitatively understand their nature. We next depict the associated phase portrait under the galactic potential in Fig.~\ref{PP1} to analyze the relevant critical points. We plot the phase topology using the black-hole potential $\Phi^{\mathrm{BH}}_i$ in the left panel, and the corresponding topology for the galactic potential $\Phi^{\mathrm{Gal}}i$ in the right panel, with the galactic parameter set to $\Upsilon_B = 100$. Throughout, the thermodynamics is governed by an adiabatic equation of state. The figure illustrates the variation of flow topologies by plotting the Mach number $M$ along the vertical axis and the radial distance in logarithmic scale ($\log_{10} r$) along the horizontal axis. For the selected set of parameters, the accretion flow exhibits multi-transonic behaviour, characterized by the presence of the inner ($I$), middle ($M$), and outer ($O$) sonic points. This configuration also allows for the formation of a stationary shock, which appears as a discontinuous jump between the supersonic and subsonic branches of the flow. The Rankine–Hugoniot shock conditions (equation~\eqref{RH3}) connect these two branches at the points $CC1$ and $DD1$. Important differences arise when comparing the positions of the sonic points and shock locations between the two panels. Most notably, the outer sonic point $O$ in the case with the galactic potential occurs at a much larger radial distance compared to the case with only the black-hole potential. Across both panels, the values of the inner sonic point $I$ remain nearly unchanged, indicating that the innermost transonic behaviour is dominated by the central black-hole potential, as expected. When the galactic potential is taken into account, we also see a discernible outward movement in the outer sonic point $O$. The image also demonstrates that $\Phi^{\mathrm{Gal}}_i$ has a considerable impact on the outer shock location, shifting inward in all four cases and indicating enhanced compression in the post-shock region. Moreover, the shock-strength parameter $CC1$ increases appreciably in the presence of $\Phi^{\mathrm{Gal}}_i$, whereas $DD1$ remains largely unchanged. The increase in shock strength implies a higher density and a thicker flow in the post-shock region, reflecting the influence of the additional galactic components. This also suggests that parameter estimation requires much greater precision for such galaxies, since the galactic potential substantially reshapes the outer transonic topology and modifies both the nature of the discontinuity and the post-shock properties.

\begin{table}
\renewcommand{\arraystretch}{1.4}
\setlength{\tabcolsep}{3pt}
  \centering
  \begin{tabular}{|l| l| r| r| r| r| r|}
    \toprule
    \textbf{Panel} & \textbf{Potential} & $\boldsymbol{I}(r_g)$ & $\boldsymbol{M}(r_g)$ & $\boldsymbol{O}(r_g)$ & $\boldsymbol{CC1}(r_g)$ & $\boldsymbol{DD1}(r_g)$ \\
    \midrule
    (a) & $\Phi^{\mathrm{BH}}_1$  & 2.23 & 8.70  & 289.88 & 151.73 & 2.43  \\
        & $\Phi^{\mathrm{Gal}}_1$ & 2.32 & 7.58  & 973.04 & 26.88  & 3.17  \\
    \midrule
    (b) & $\Phi^{\mathrm{BH}}_2$  & 2.08 & 8.64  & 290.44 & 123.15 & 2.38  \\
        & $\Phi^{\mathrm{Gal}}_2$ & 2.18 & 7.38  & 981.18 & 25.45  & 3.10  \\
    \midrule
    (c) & $\Phi^{\mathrm{BH}}_3$  & 1.17 & 4.27  & 295.40 & 103.05 & 1.25  \\
        & $\Phi^{\mathrm{Gal}}_3$ & 1.17 & 3.39  & 979.01 & 28.01  & 1.38  \\
    \midrule
    (d) & $\Phi^{\mathrm{BH}}_4$  & 1.50 & 5.73  & 293.57 & 98.55  & 1.66  \\
        & $\Phi^{\mathrm{Gal}}_4$ & 1.52 & 5.21  & 977.02 & 29.36  & 1.90  \\
    \bottomrule
  \end{tabular}
 \caption{Computed quantities for panels (a)--(d) showing the variation of $I$, $M$, $O$, $CC1$, and $DD1$ for both $\Phi^{\mathrm{BH}}_i$ and $\Phi^{\mathrm{Gal}}_i$ as shown in figure \ref{PP1}.}
   \label{phase-tab}
\end{table}

To study the flow paths and phase portraits, we numerically integrate the governing equations starting from the critical values of the velocity and its radial slope. This method gives a clear picture of the possible flow types—smooth flows or flows with shocks—based on the chosen boundary conditions and parameters. However, these plots alone cannot fully explain how the flow behaves. Numerical integration always gives a trajectory, but it does not tell us whether the related critical point is physically meaningful. Therefore, in the next subsection, we study a detailed dynamical systems analysis, to understand the true qualitative nature of the flow and describe the properties of the critical points obtained in the presence of the combined galactic potential.

\subsection{Nature of critical points}
\begin{figure*}
  \centering
  \begin{subfigure}[b]{0.8\linewidth}
    \includegraphics[width=0.50\linewidth,clip]{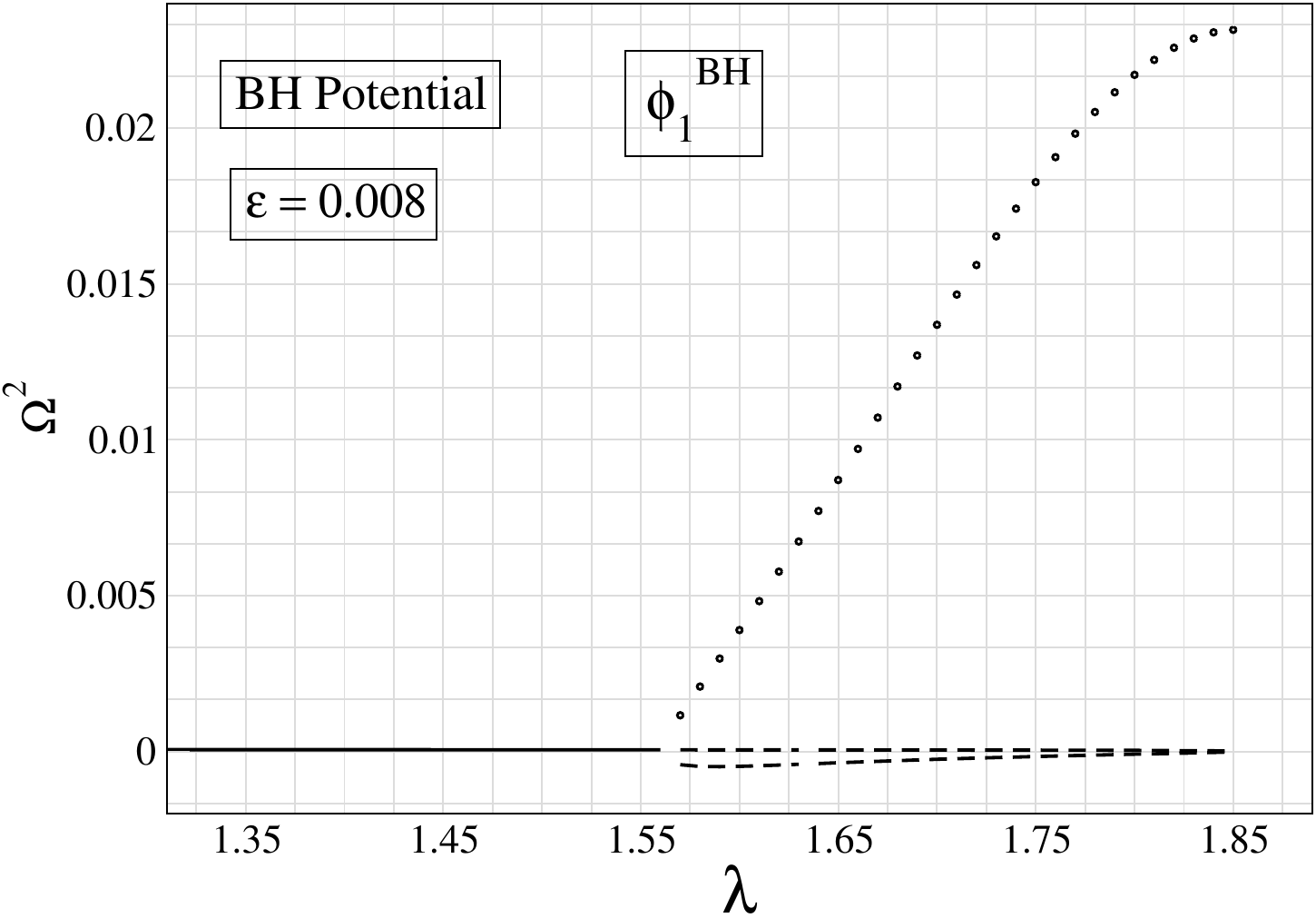}
    \includegraphics[width=0.50\linewidth,clip]{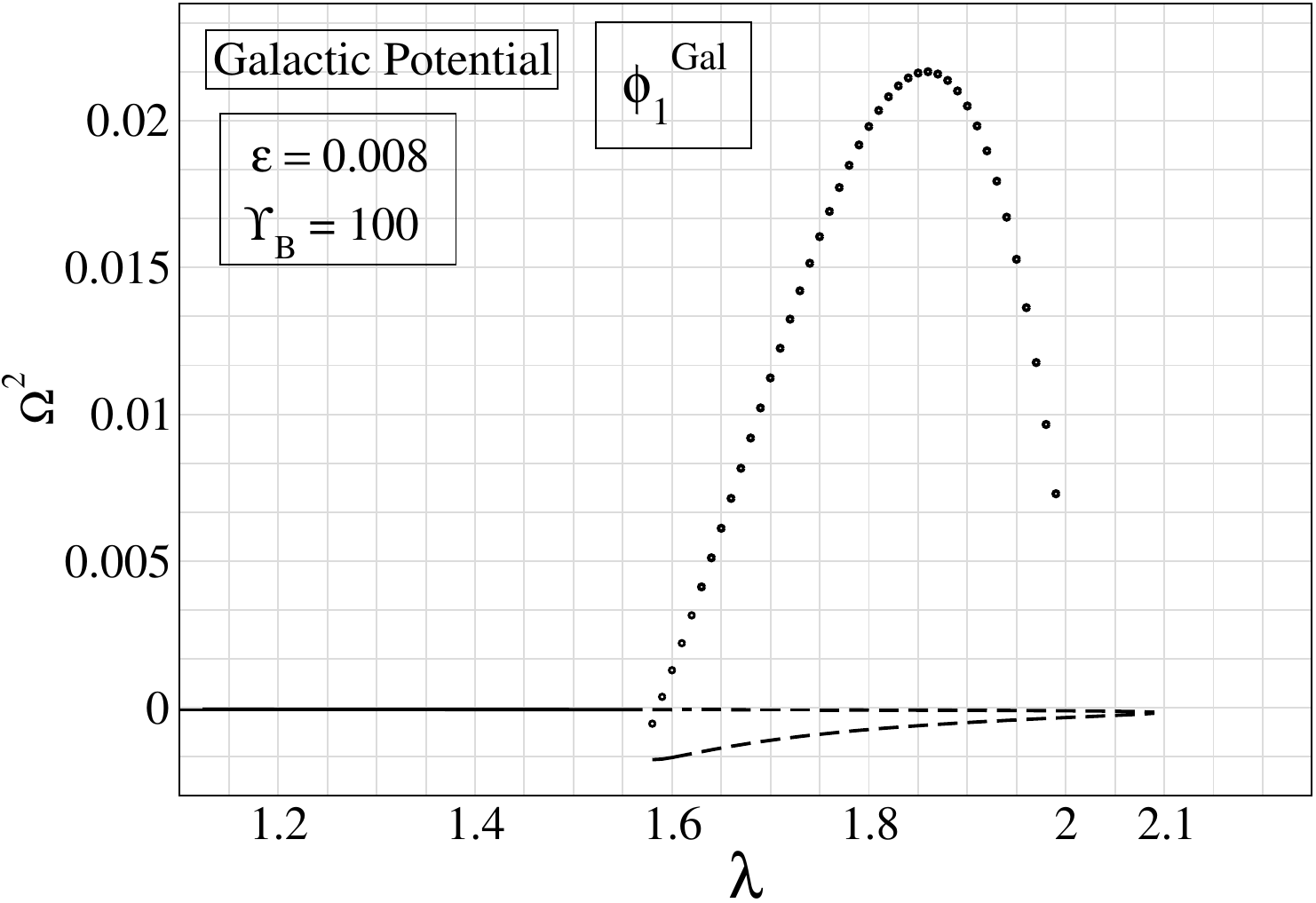}
    \caption{}
    \label{bhgala}
  \end{subfigure}
  \begin{subfigure}[b]{0.8\linewidth}
    \includegraphics[width=0.50\linewidth,clip]{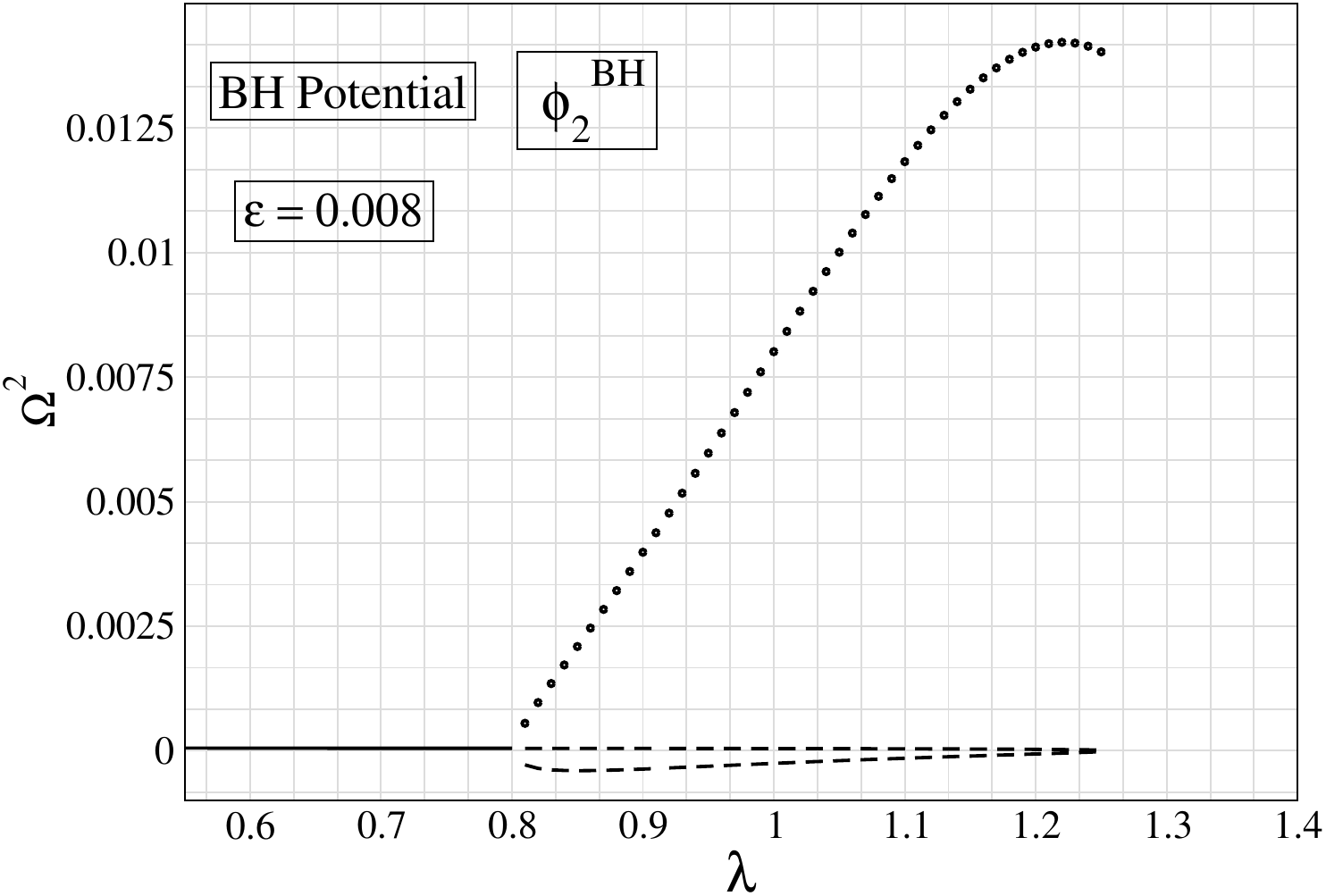}
    \includegraphics[width=0.50\linewidth,clip]{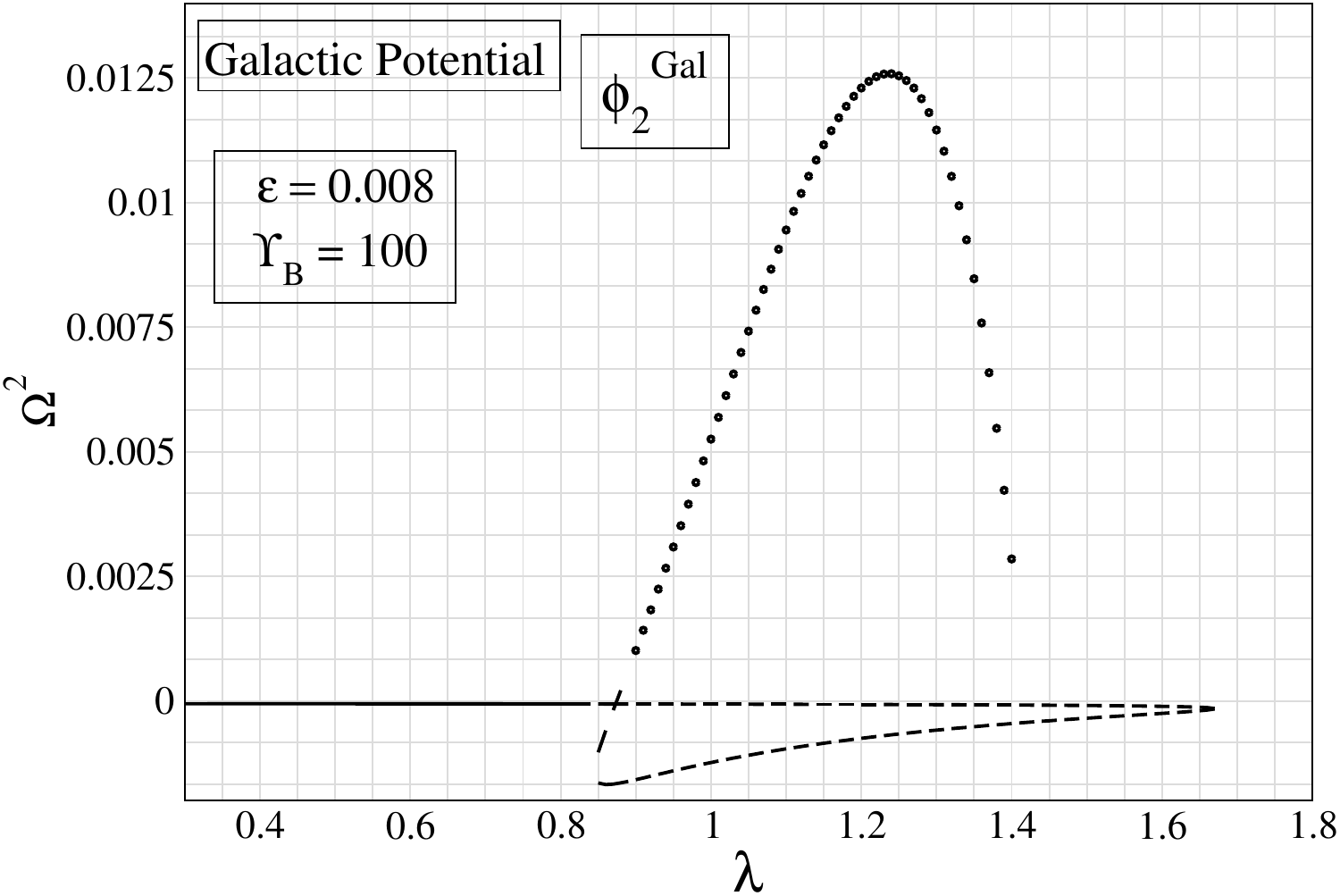}
    \caption{}
    \label{bhgalb}
  \end{subfigure}

  \begin{subfigure}[b]{0.8\linewidth}
    \includegraphics[width=0.50\linewidth,clip]{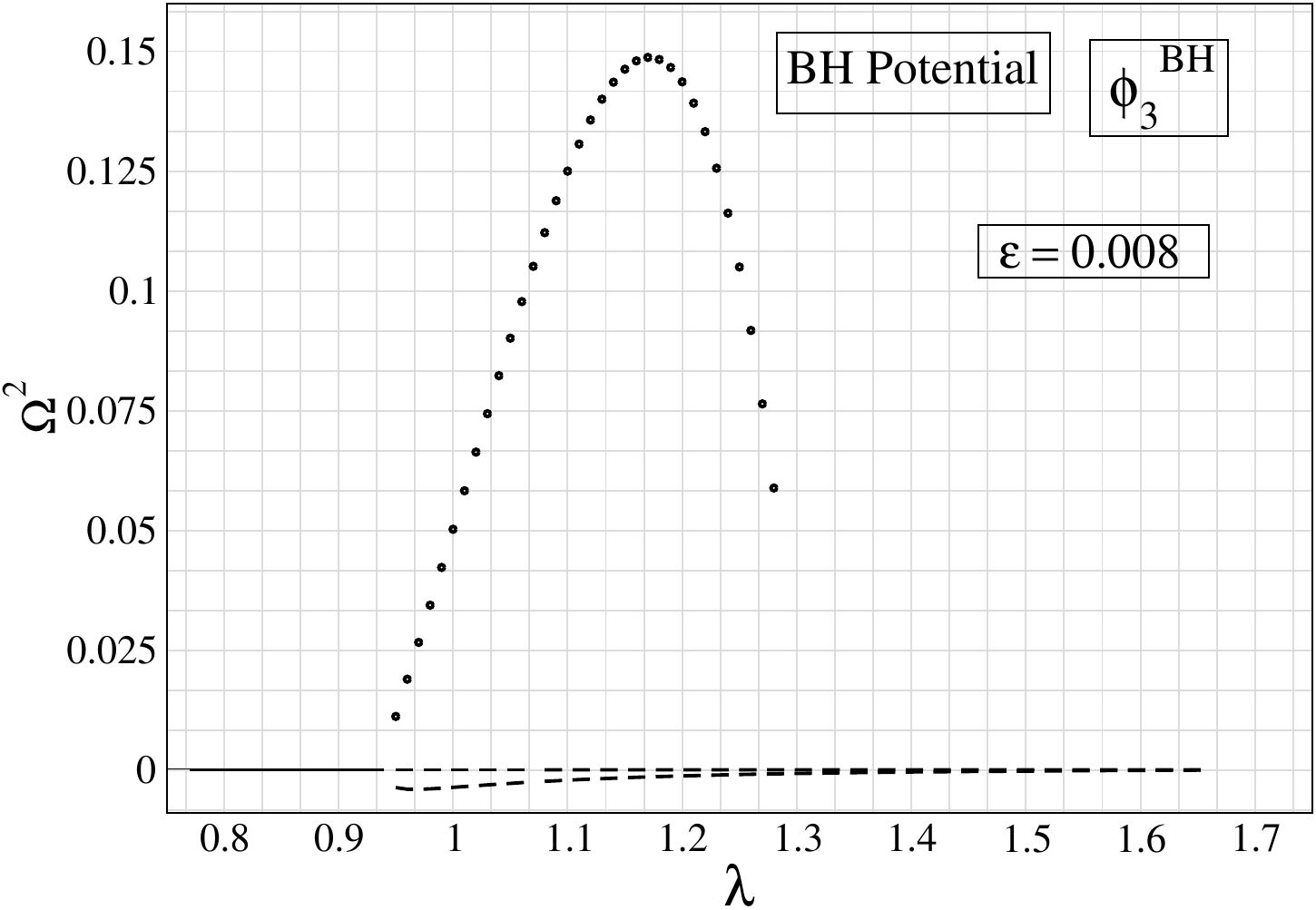}
    \includegraphics[width=0.50\linewidth,clip]{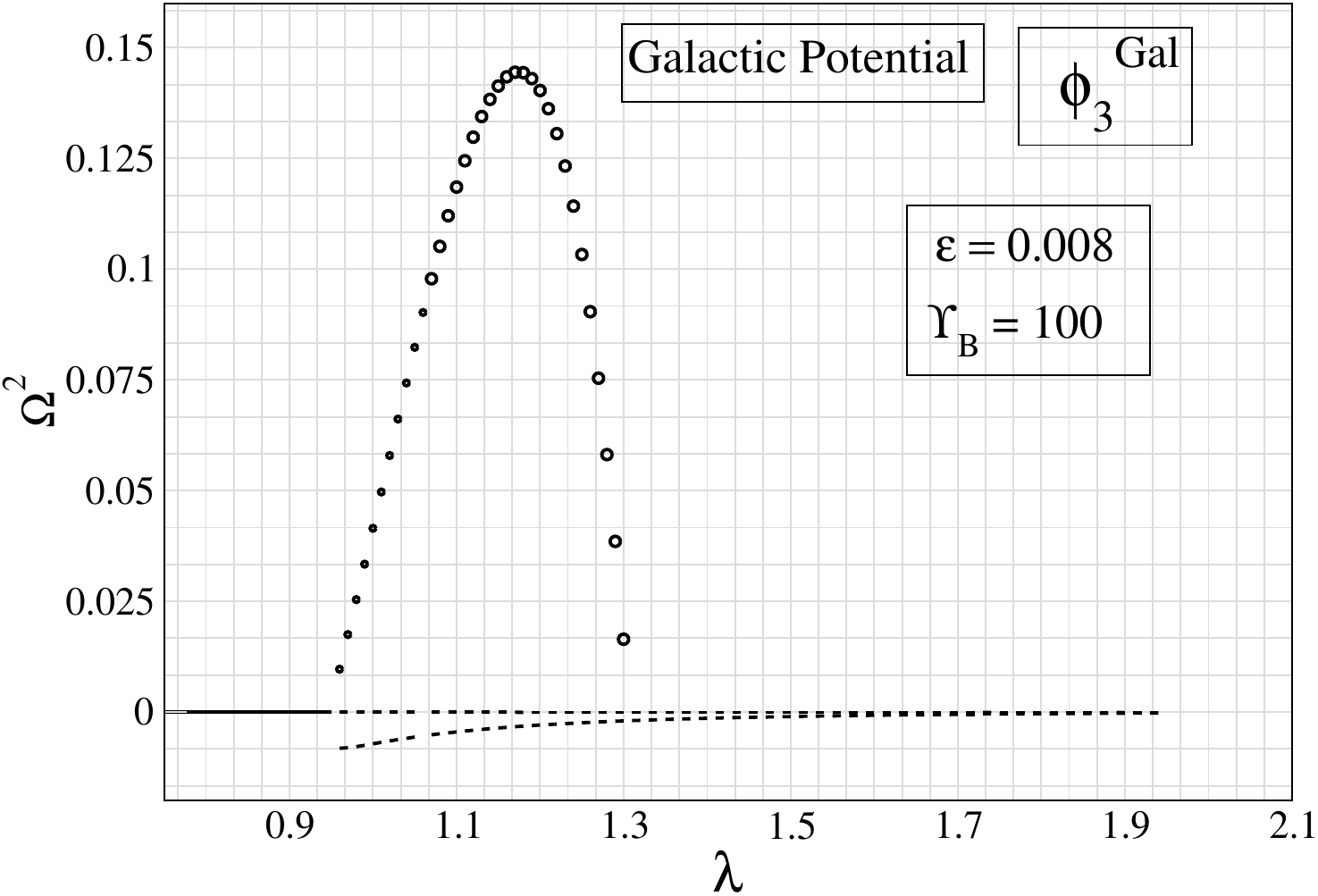}
    \caption{}
    \label{bhgalc}
  \end{subfigure}
  \begin{subfigure}[b]{0.8\linewidth}
   \includegraphics[width=0.50\linewidth,clip]{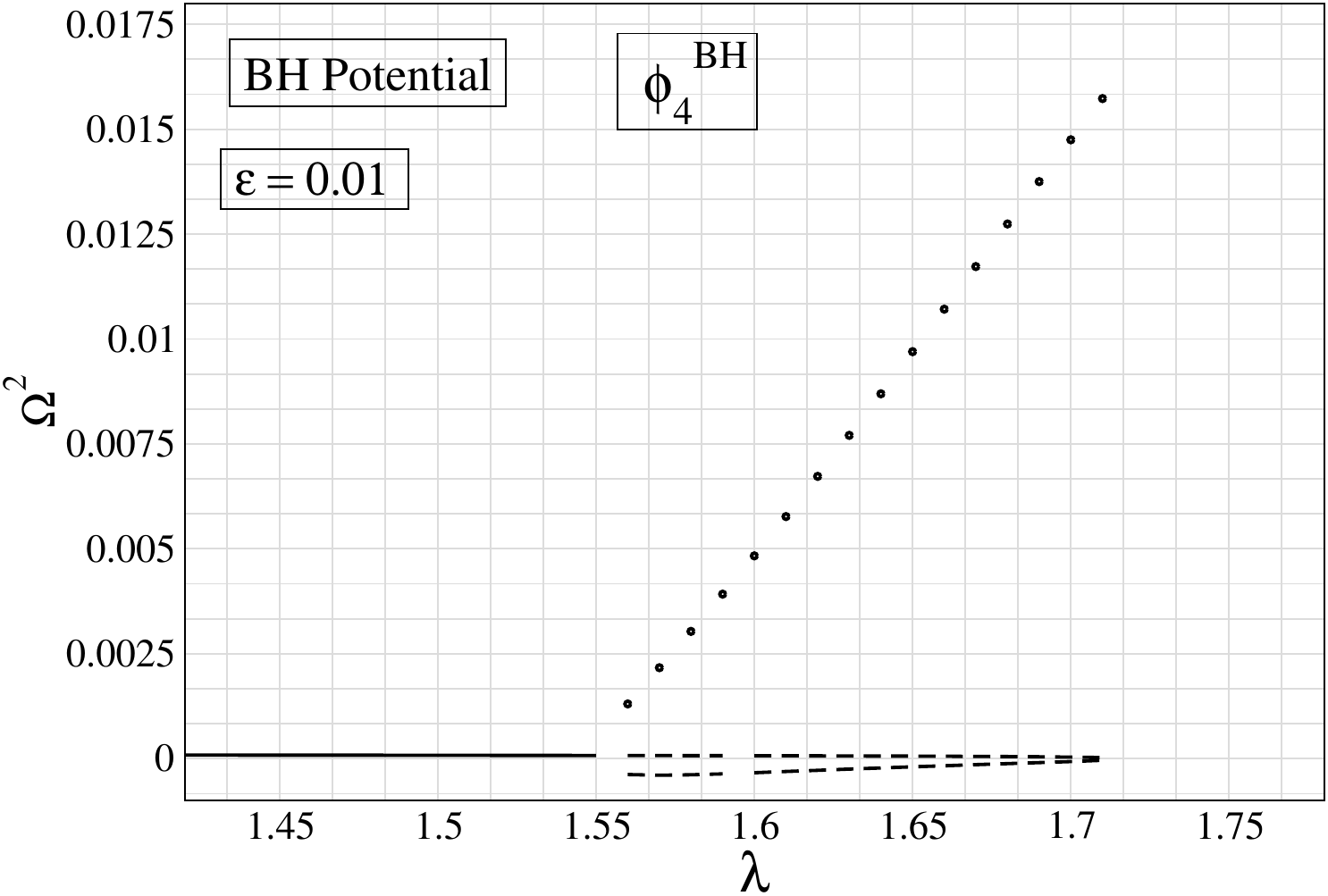}
    \includegraphics[width=0.50\linewidth,clip]{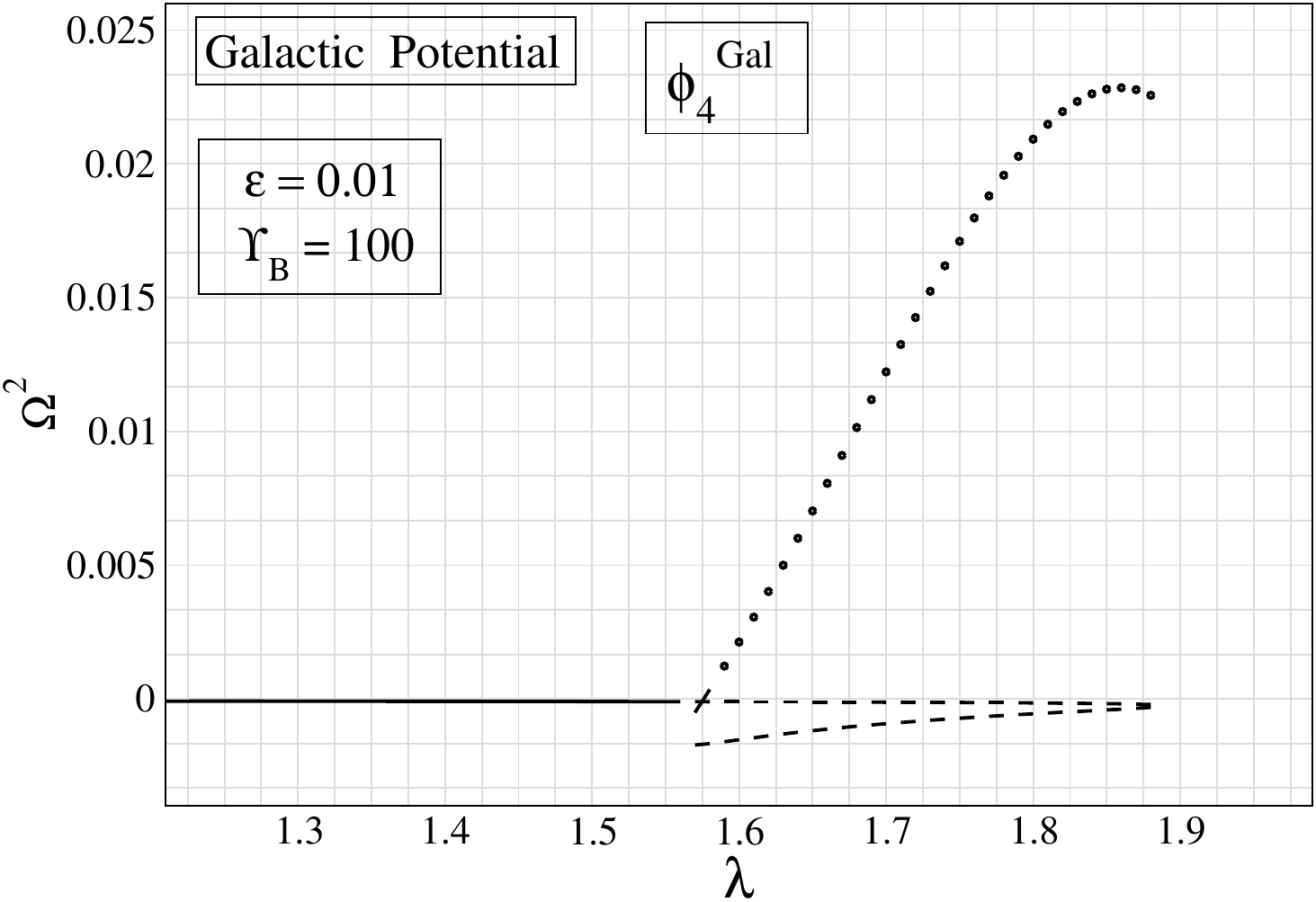}
    \caption{}
    \label{bhgald}
  \end{subfigure}
   \caption{In this figure, we present the nature of critical points under the BH pseudo potentials $\Phi^{\mathrm{BH}}_i$ in left panel and galactic potential $\Phi^{\mathrm{Gal}}_i$ in right panel with galactic parameter $\Upsilon_B = 100$. A saddle point is indicated by $\Omega^2 > 0$ in all subplots, while a center-type point is indicated by $\Omega^2 < 0$.  A typical sequence is depicted in each subplot: a single saddle point at low $\lambda$; the emergence of a centre and an additional saddle via a saddle--center bifurcation; and, at higher $\lambda$, the annihilation of the centre with the outer saddle through a reverse saddle--center bifurcation, leaving one outer saddle to persist.  A horizontal line is used to plot the lone saddle.  The creation of a homoclinic loop and its homoclinic connection at greater $\lambda$ are indicated by $-$ markers, whilst the evolution of $\Omega^2$ for all three critical points is displayed by $\circ$ markers.  The galactic environment has a noticeable impact on the critical points overall behaviour.}
\label{nature-eigen-CHCOVE}
\end{figure*}
However, in order to provide some quantitative information on the specific features of each critical point, the behaviour of the eigenvalues of the stability matrix associated with each critical point must be investigated.  For potentials $\Phi^{\mathrm{BH}}_i$ and $\Phi^{\mathrm{Gal}}_i$ with galactic parameter $\Upsilon_B = 100$, the figure \ref{nature-eigen-CHCOVE} is obtained.  Plotting this requires going back to equations \eqref{OVE}, which give a dependency of $\Omega^2$ on the crucial point's coordinates.  When $\Omega^2$ is negative, it denotes a centre point where the flow travels in closed circular paths without connecting regions like infinity and the event horizon; when $\Omega^2$ is positive, it is a saddle point, which is unstable but permits the flow to transition through it along specific directions. The system initially has only one saddle point (shown as a $ -$ sign in figure \ref{nature-eigen-CHCOVE} ) as the angular momentum $\lambda$ increases from a sufficiently low value.  A saddle-center bifurcation then takes place at a critical $\lambda$, creating two new critical points: an inner saddle point at $r_{in}$ and a centre point at a radius called $r_{mid}$, while the original saddle point moves outward to become the outer saddle at $r_{out}$.  The behaviour of $\Omega^2$ for the three critical points is illustrated by $\circ$ sign curves in the figure \ref{nature-eigen-CHCOVE} , and throughout this interval of $\lambda$, the connection $\dot{M}_{in} > \dot{M}_{out}$ (see equation \eqref{ent}) holds true.
Both polytropic and isothermal flows exhibit the same general behaviour; the main differences are in the parameters ($\dot{M}$ for polytropic, $C$ for isothermal).  In the wedge-shaped region that corresponds to multicritical points, a particularly intriguing and crucial scenario emerges along the boundary between two subregions of the parameter space.  In contrast to the homoclinic loop, where a saddle connects back to itself, in this case the condition $\dot{M}_{in} = \dot{M}_{out}$ (or $C_{in} = C_{out}$ see \eqref{iso1}) is satisfied, resulting in a situation where the two saddle points are directly connected by their separatrices, forming a heteroclinic connection. The homoclinic connection belongs to the inner saddle just below this boundary value of $\lambda$, while the homoclinic loop moves to the outer saddle point just above it.  From the viewpoint of general dynamical systems, this shift indicates the particular kind of bifurcation characterised as a heteroclinic bifurcation, in which the flow trajectories' topology undergoes a fundamental change.  Despite the fact that the Mach number is equal to one at both saddle and centre points mathematically, the physically significant accretion flow—the transonic solution that connects infinity to the black hole's event horizon—can only pass through saddle-type critical points because only saddle points allow the required smooth transition from subsonic to supersonic flow.For this purpose, in astrophysical accretion flows, only saddle-type sites are regarded as real sonic points.The galactic environment has a considerably influence impact on the flow topology, , as evidenced by the overall character and evolution of the critical points consequently changing in response to it.
\subsection{Influence of galactic parameter on shock location and multi-transonic accretion solution -Shock-strength and shock-induced flow variables}
\begin{figure*}
 \centerline{\includegraphics[width=0.9\linewidth,clip]{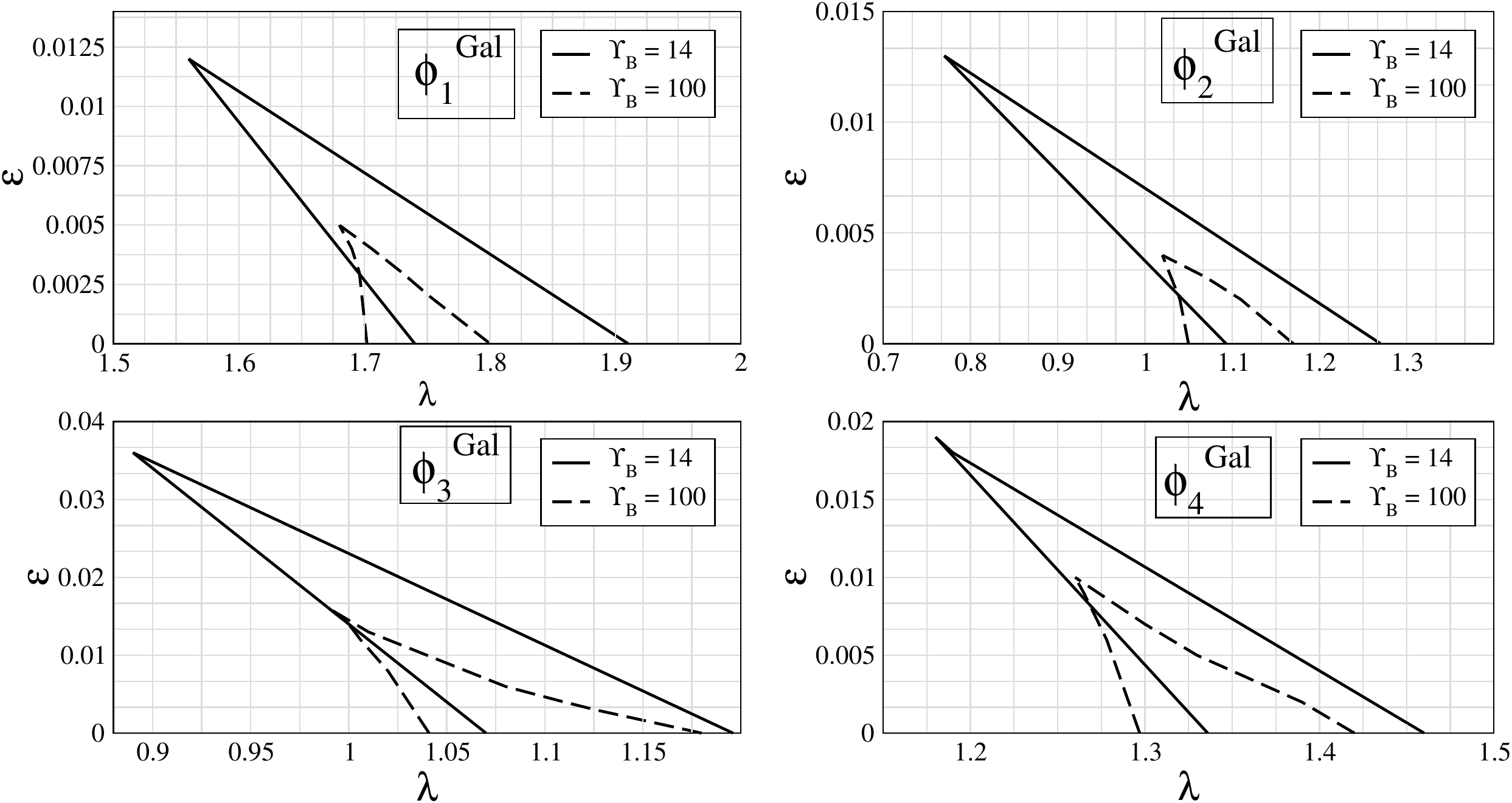}}
  \caption{In this figure we compare the shock-permitting regions in the parameter space ($\mathcal{E}$--$\lambda$) with the galactic potential $\Phi^{\mathrm{Gal}}_i$ embedded in four pseudo Schwarzschild potentials. We assume  two values of the galactic parameter given as, solid lines correspond to $\Upsilon_B = 14$, while dashed lines correspond to $\Upsilon_B = 100$. The figure highlights the influence of $\Upsilon_B$ on the size and location of the shock-permitting regions, illustrating how the inclusion of a stronger galactic potential affects the conditions for shock formation in a multi-component galactic environment. The comparison demonstrates that increasing the galactic parameter $\Upsilon_B$ reduces and shifts the shock-permitting region in the $\mathcal{E}$--$\lambda$ plane. The four panels also compare the four potentials, for which we see that the ($\mathcal{E}$--$\lambda$) space is distinctly different and with the increases in $\Upsilon_B$ from 14 to 100 yields almost less than half of the available space. This indicates higher mass distribution of the galaxy suppress the available region for shock formation.
  }
\label{threeshockCHCOVE}
\end{figure*}
\begin{figure*}
 \centerline{\includegraphics[width=0.85\linewidth,clip]{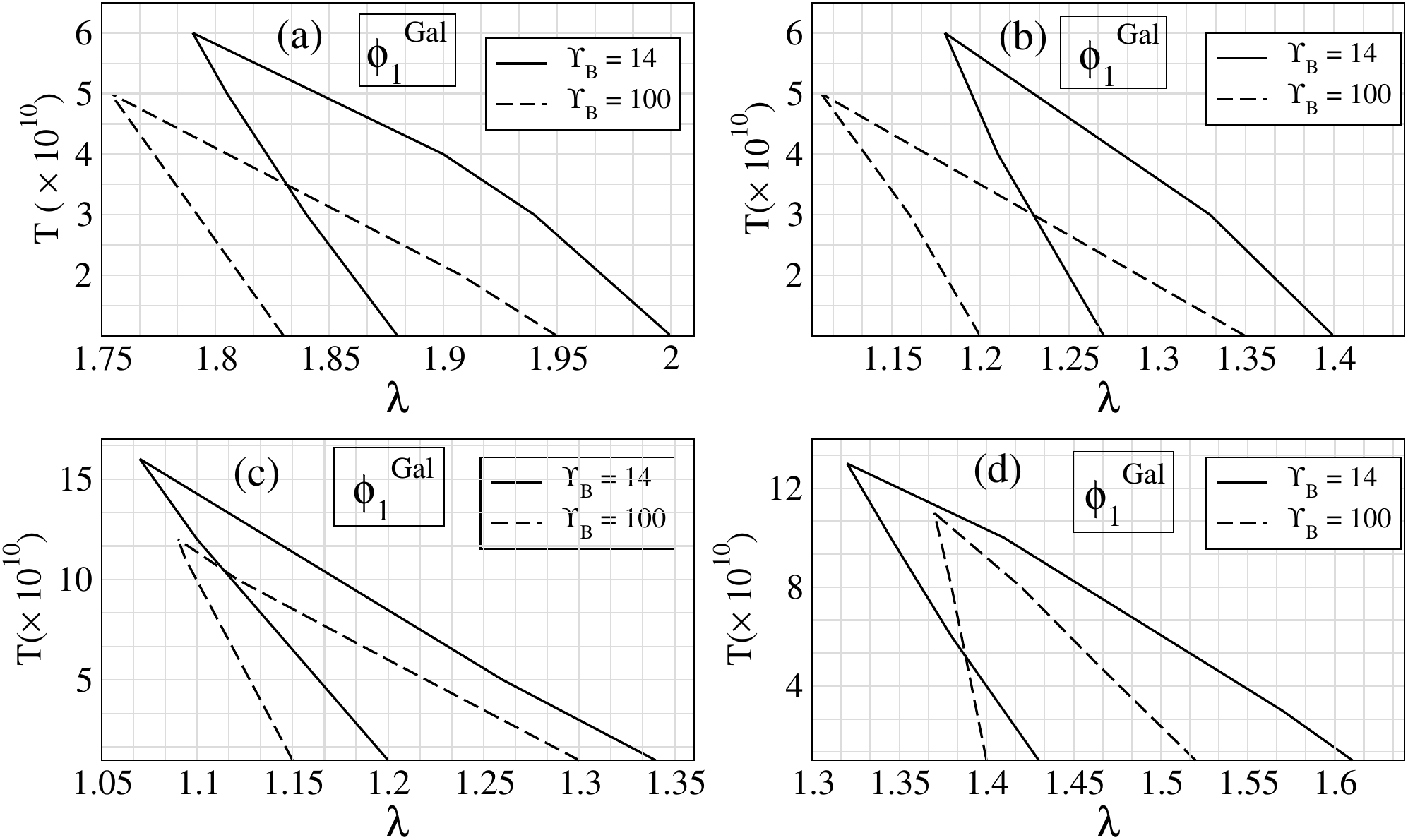}}
   \caption{In this figure, we compare the shock-permitting regions in the temperature–angular momentum ($T$--$\lambda$) parameter space  with the galactic potential $\Phi^{\mathrm{Gal}}_i$ with the chosen pseudo Schwarzschild potential. We assume  two values of the galactic parameter given as, solid lines correspond to $\Upsilon_B = 14$, while dashed lines correspond to $\Upsilon_B = 100$. The figure highlights the comparative behavior and the influence of $\Upsilon_B$ on shock formation in a multi-component galactic potential. Increasing $\Upsilon_B$ for all the four cases of pseudo-Schwarzschild potential reduces and shifts the shock-permitting region, indicating that a stronger galactic potential stabilizes the outer flow and narrows the range of temperature and angular momentum for which shocks can occur. This emphasizes an essential aspect of the multi-component galactic environment in shaping the outer flow structure and shock characteristics around a central black hole.
   }
\label{shock-ISO}
\end{figure*}
Shock formation in accretion must adhere to the Rankine-Hugoniot requirements and conserve energy in the case of an adiabatic equation of state, as explained in \S \ref{shok-F}. In figures \ref{threeshockCHCOVE} and \ref{shock-ISO}, we analyze and compare these shock-permitting regions, considering an adiabatic and isothermal flow we presents a comparison of shock-permitting regions for the all the pseudo Schwarzschild potential in the specific energy–angular momentum ($\mathcal{E}$--$\lambda$) and  temperature–angular momentum ($T$--$\lambda$) parameter space under the galactic potential $\Phi^{\mathrm{Gal}}_i$. Two values of the galactic parameter, $\Upsilon_B = 14$ (solid lines) and $\Upsilon_B = 100$ (dashed lines), are considered to examine the effect of galactic strength on shock formation. Fig \ref{threeshockCHCOVE} \& fig \ref{shock-ISO} shows that for $\Upsilon_B=100$ size of the shock-permitting region and shifts it toward lower angular momentum and energy values for adiabatic case and temperature for the isothermal case for all the four pseudo Schwarzschild potential. These finding indicates that a stronger galactic potential makes it harder for the flow to develop a shock unless the flow is already more tightly bound and having lower angular momentum. Physically, the galaxy’s gravity stabilizes the outer parts of the flow and suppresses shock formation, allowing shocks to appear only in low-energy/low-temperature, low-angular-momentum configurations.

\begin{table}
\centering
\renewcommand{\arraystretch}{1.2}
\setlength{\tabcolsep}{4pt}
\begin{tabular}{|c|c|c|c|c|c|c|c|c|}
\hline
Potential
& \multicolumn{2}{c|}{$\Phi^{\mathrm{Gal}}_{1}$}
& \multicolumn{2}{c|}{$\Phi^{\mathrm{Gal}}_{2}$}
& \multicolumn{2}{c|}{$\Phi^{\mathrm{Gal}}_{3}$}
& \multicolumn{2}{c|}{$\Phi^{\mathrm{Gal}}_{4}$} \\
\hline
& $\mathcal{E}$ & $\lambda$
& $\mathcal{E}$ & $\lambda$
& $\mathcal{E}$ & $\lambda$
& $\mathcal{E}$ & $\lambda$ \\
\hline
$\Upsilon_{B=14}$  &0.0124  &1.74 &0.013&1.05&0.037&1.04&0.019&1.30  \\
\hline
$\Upsilon_{B=100}$ &0.0051  & 1.70 &0.004&1.10&0.015&1.06&0.01&1.33  \\
\hline
\end{tabular}
\caption{Energy ($\mathcal{E}$) and specific angular momentum ($\lambda$) corresponding to different galactic potentials see figure \ref{threeshockCHCOVE}.}
\label{tab:E_lambda_potentials}
\end{table}

\subsection{Shock-strength and shock-induced flow variables}
\begin{figure*}
  \centering
  \begin{subfigure}[b]{0.49\linewidth}
    \includegraphics[width=\linewidth,clip]{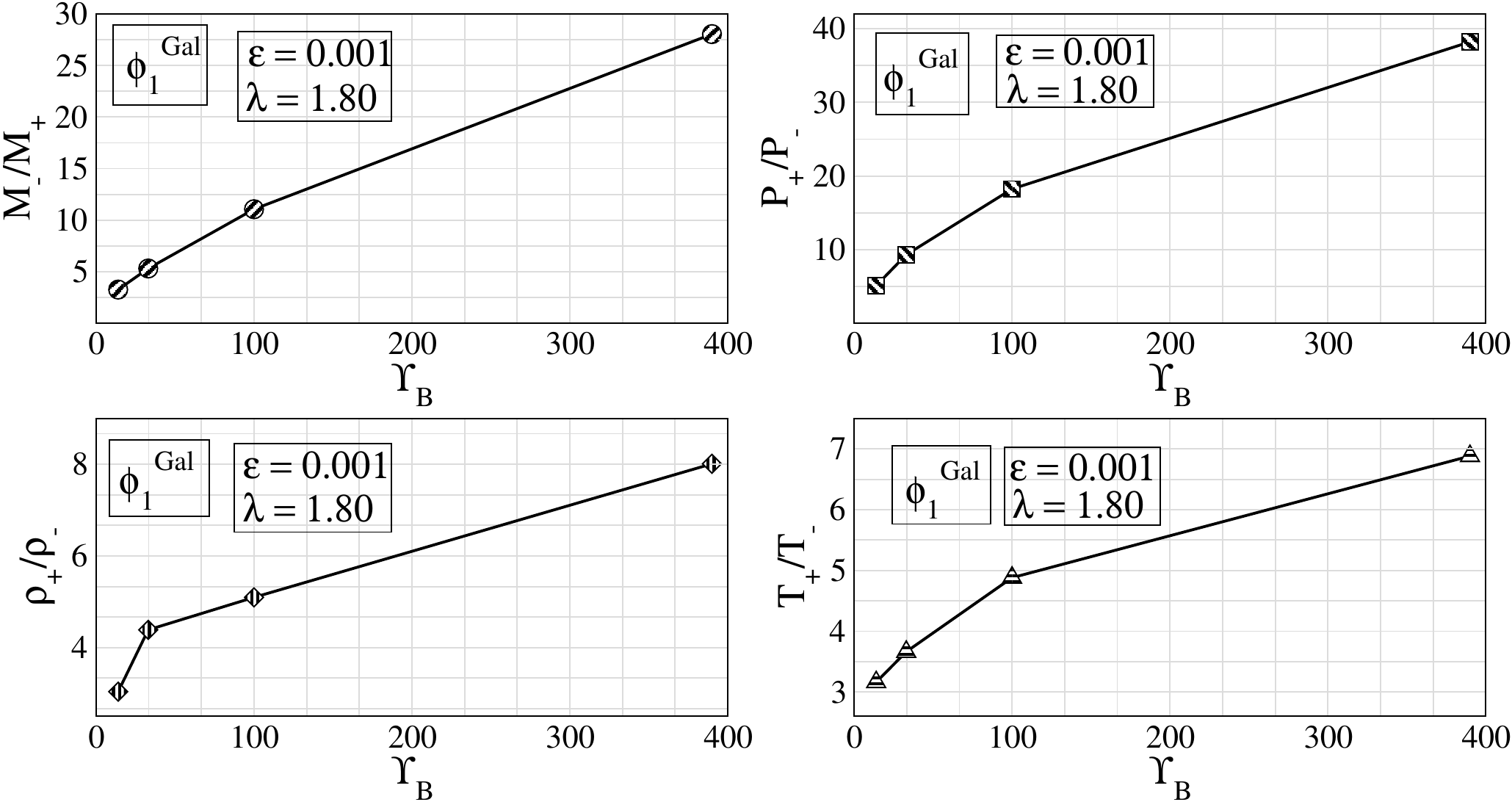}
    \caption{}
    \label{MPTa}
  \end{subfigure}
  \begin{subfigure}[b]{0.49\linewidth}
    \includegraphics[width=\linewidth,clip]{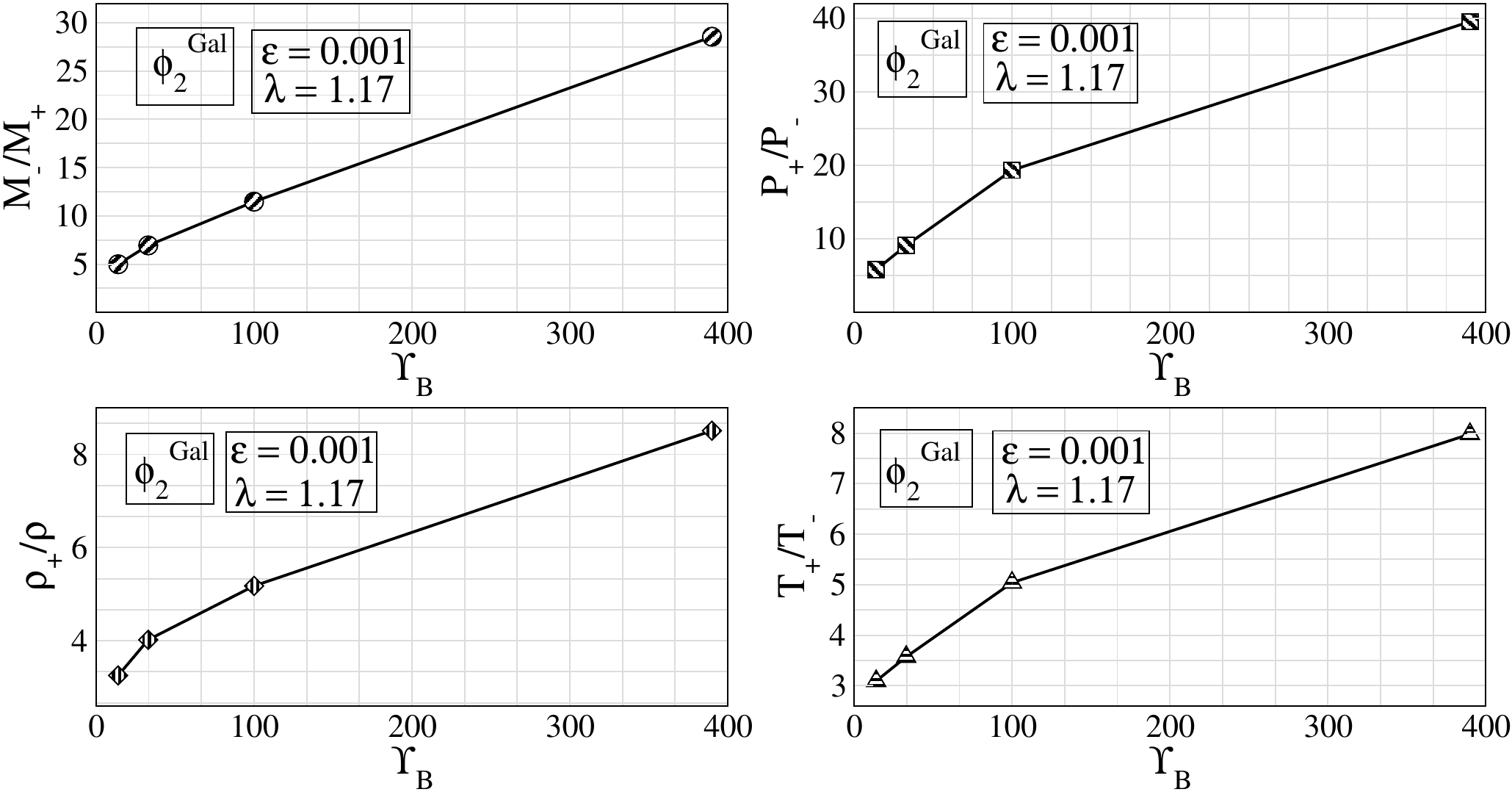}
    \caption{}
    \label{MPTb}
  \end{subfigure}

  \begin{subfigure}[b]{0.49\linewidth}
    \includegraphics[width=\linewidth,clip]{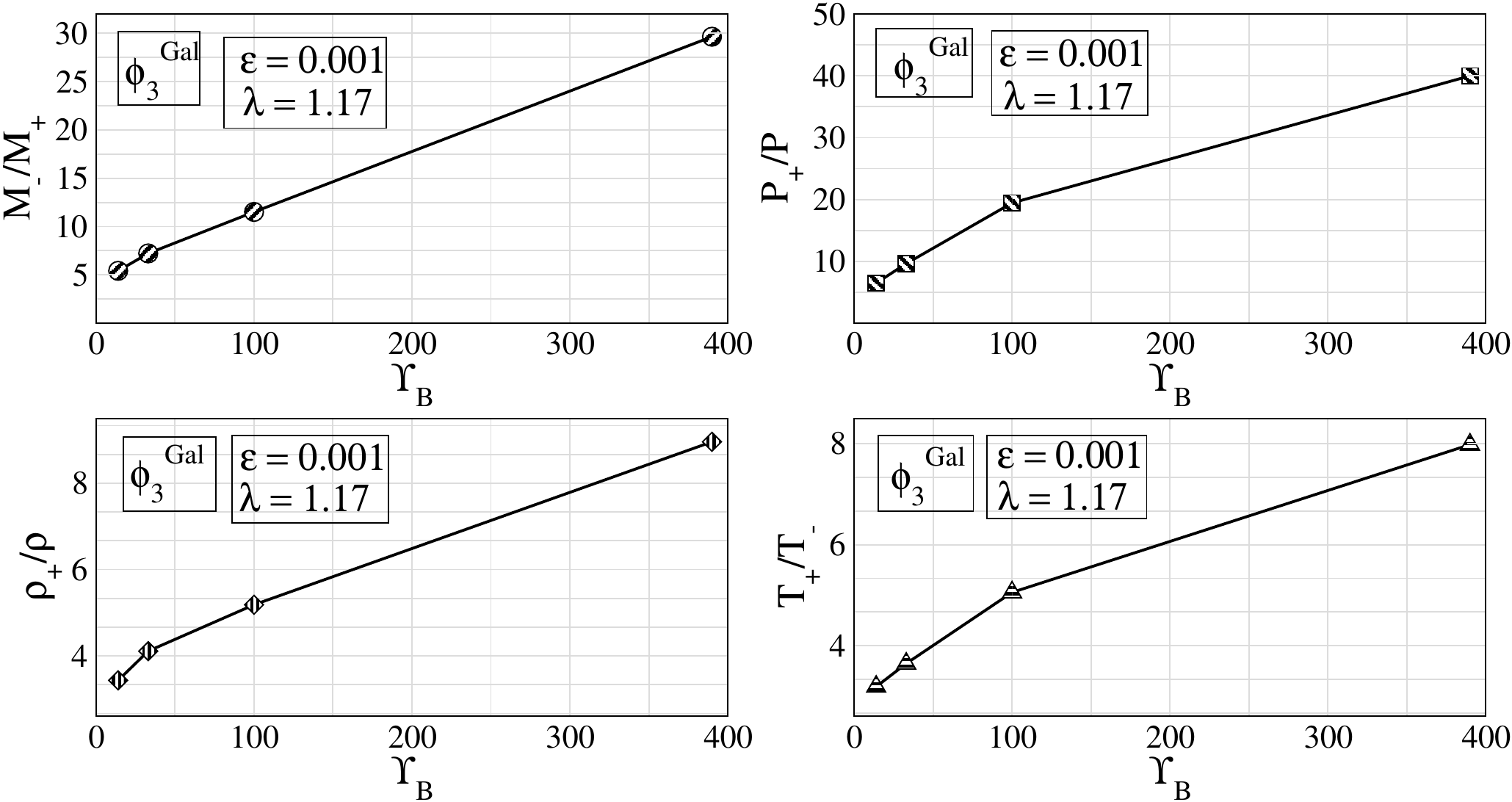}
    \caption{}
    \label{MPTc}
  \end{subfigure}
  \begin{subfigure}[b]{0.49\linewidth}
    \includegraphics[width=\linewidth,clip]{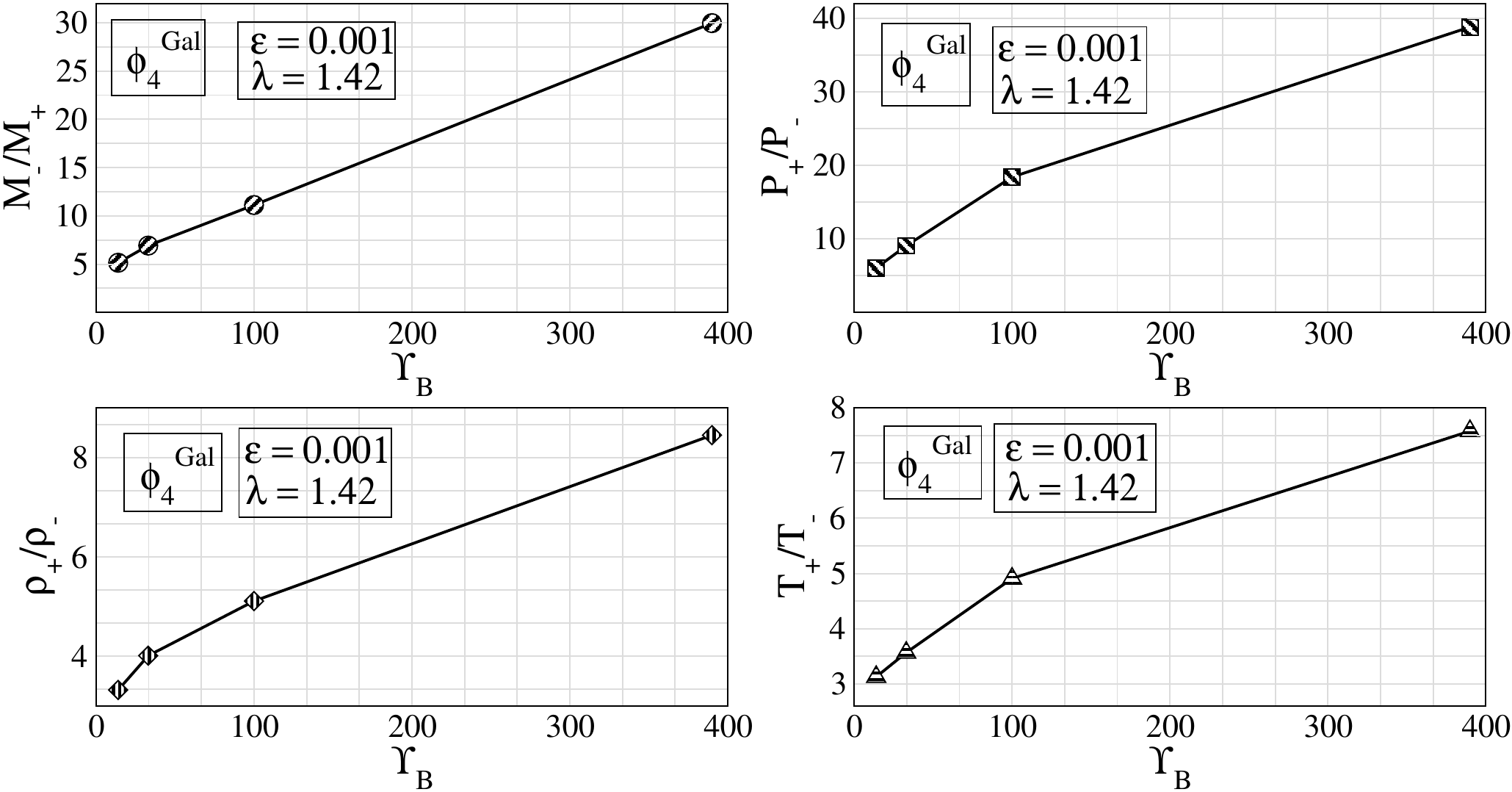}
    \caption{}
    \label{MPTd}
  \end{subfigure}
  \caption{ In this figure we present the variation of shock strength  for the galactic parameter: $\Upsilon_{B} = 14$, $\Upsilon_{B} = 33$, $\Upsilon_{B} = 100$, and $\Upsilon_{B} = 390$ considering four pseudo Schwarzschild potential. The results are obtained  for  $\mathcal{E}$, $\lambda$] = [0.001, 1.80] in panel (a), for  $\mathcal{E}$, $\lambda$] = [0.001, 1.17] in panel (b), for  $\mathcal{E}$, $\lambda$] = [0.001, 1.17] in panel (c) and for  $\mathcal{E}$, $\lambda$] = [0.001, 1.42] in panel (d) respectively. Each panel shows the variation of the shock strength in terms of the Mach number ratio ($M_{-} / M_{+}$), pressure ratio across the shock ($P_{+} / P_{-}$) and also illustrates the variation of the post-shock to pre-shock temperature ratio ($T_{+} / T_{-}$), and panel (d) displays the compression ratio ($\rho_{+} / \rho_{-}$) as a function of increasing galactic potential. These panels collectively highlight how different galactic environments, characterized by increasing $\Upsilon_{B}$ alongwith the chosen pseudo potential, influence the strength and thermodynamic properties of shocks in the transonic accretion flow.}
  \label{MPT-phi}
\end{figure*}
The perspective of shock strength, which quantifies the difference between the pre-shock and post-shock flow states, is introduced in this section to explore the properties of shocks in the presence of a galactic potential. The amount of compression, entropy generation, and post-shock pressure across the shock front are all influenced by the shock strength. Strong shocks are linked to significant entropy gradients, while weak shocks are connected with small compression ratios. Physically, the shock strength defines the thermodynamic and dynamical character of the post-shock region under the galactic potential by measuring the efficiency of compression and the conversion of kinetic energy into thermal energy, thereby quantifying how abruptly and irreversibly the accretion flow is restructured at the shock front.
We plot  figure \ref{MPT-phi} for the four pseudo Schwarzschild potential to show how the strength of shock waves changes when the galactic parameter $\Upsilon_{B}$) increases. The values of $\Upsilon_{B}$ used are 14, 33, 100, and 390. The results are shown for  $\mathcal{E}$, $\lambda$] = [0.001, 1.80] in panel (a), for  $\mathcal{E}$, $\lambda$] = [0.001, 1.17] in panel (b), for  $\mathcal{E}$, $\lambda$] = [0.001, 1.17] in panel (c) and for  $\mathcal{E}$, $\lambda$] = [0.001, 1.42] in panel (d) respectively. Each panel , the shock strength is shown using the Mach number ratio ($M_{-} / M_{+}$), which tells us how much the speed of the flow changes across the shock. Panel (b) shows the pressure ratio ($P_{+} / P_{-}$), indicating how much the pressure jumps due to the shock. Panel (c) shows the temperature ratio ($T_{+} / T_{-}$), which reflects how hot the flow becomes after the shock. Panel (d) presents the compression ratio ($\rho_{+} / \rho_{-}$), showing how much denser the flow gets after passing through the shock. From these figures we can see that shock waves in the accretion flow behave differently depending on the galaxy’s mass distribution. As $\Upsilon_{B}$ becomes larger, the shock becomes stronger and its effect on the flow becomes more noticeable.

\begin{figure*}
\centerline{\includegraphics[width=0.7\linewidth,clip]{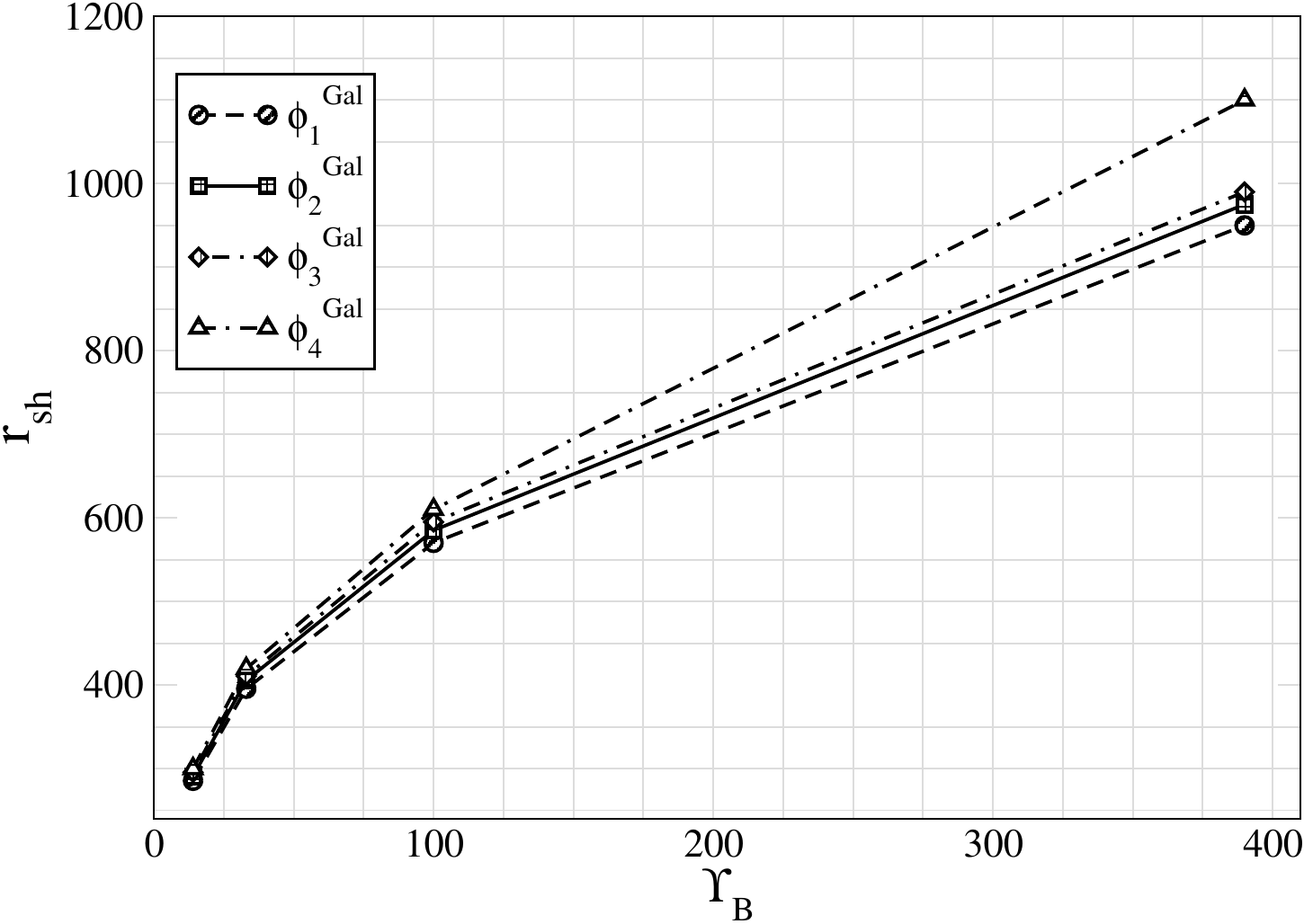}}
  \caption{This figure present the variation of shock location for the galactic potential $\Phi^{\mathrm{Gal}}_i$ considering the four pseudo-Schwarzschild potential as described. We consider  $\Upsilon_{B} = 14$, $\Upsilon_{B} = 33$, $\Upsilon_{B} = 100$, and $\Upsilon_{B} = 390$ to plot this figure. The figure illustrates how the position of shock formation in axisymmetric transonic flows is influenced by the strength of the galactic potential and the associated pseudo Schwarzschild potential. An increase in $\Upsilon_{B}$, corresponding to a stronger contribution from the surrounding galactic components, generally causes the shock location to shift outward, reflecting the growing dynamical influence of the extended galactic environment on the accretion flow. We also see that at the lower galactic scale parameter range all the four potential are nearly giving same shock location, however as $\Upsilon_B$ increases the four potential distinctly shows variation in shock location.}
\label{shocklocationCHCOVE}
\end{figure*}

This is followed by an analysis of the variation of the shock location for the four pseudo-Schwarzschild potentials embedded in $\Phi^{\text{Gal}}_i$, as shown in fig \ref {shocklocationCHCOVE}.The analysis is conducted  $\Upsilon_{B} = 14$, $33$, $100$, and $390$. The findings show that the radial position of shock production with the chosen pseudo-Schwarzschild potential has been substantially affected by the surrounding galactic potential. For lower value of  $\Upsilon_{B}$, all the four potential provides almost same shock location however when $\Upsilon_{B}$ increases the shock front systematically shifts, with noticeable differences among the four potential models.

\begin{table}
\centering
\renewcommand{\arraystretch}{1.1}
\setlength{\tabcolsep}{4pt}
\begin{tabular}{|c|c|c|c|c|}
\hline
Potential
& $\Phi^{\mathrm{Gal}}_{1}$ ($r_{\mathrm{sh}}/r_g$)
& $\Phi^{\mathrm{Gal}}_{2}$ ($r_{\mathrm{sh}}/r_g$)
& $\Phi^{\mathrm{Gal}}_{3}$ ($r_{\mathrm{sh}}/r_g$)
& $\Phi^{\mathrm{Gal}}_{4}$ ($r_{\mathrm{sh}}/r_g$) \\
\hline
$\Upsilon_{B=14}$  & 290.45 & 290.82 & 295.57 & 293.79 \\
\hline
$\Upsilon_{B=33}$  & 400.45 & 405.46 & 408.60 & 406.70 \\
\hline
$\Upsilon_{B=100}$ & 549.01 & 560.64 & 595.64 & 605.04 \\
\hline
$\Upsilon_{B=390}$ & 973.03 & 981.18 & 989.00 & 1110.75 \\
\hline
\end{tabular}
\caption{This table presents the shock locations for the VE disc model in the adiabatic case under different galactic potentials. The shock position is expressed in units of the gravitational radius $r_g$. see \ref {shocklocationCHCOVE}}
\label{tab:rsh_percentage}
\end{table}

\subsection{Acoustic surface gravity}
\begin{figure*}
\centerline{\includegraphics[width=0.7\linewidth,clip]{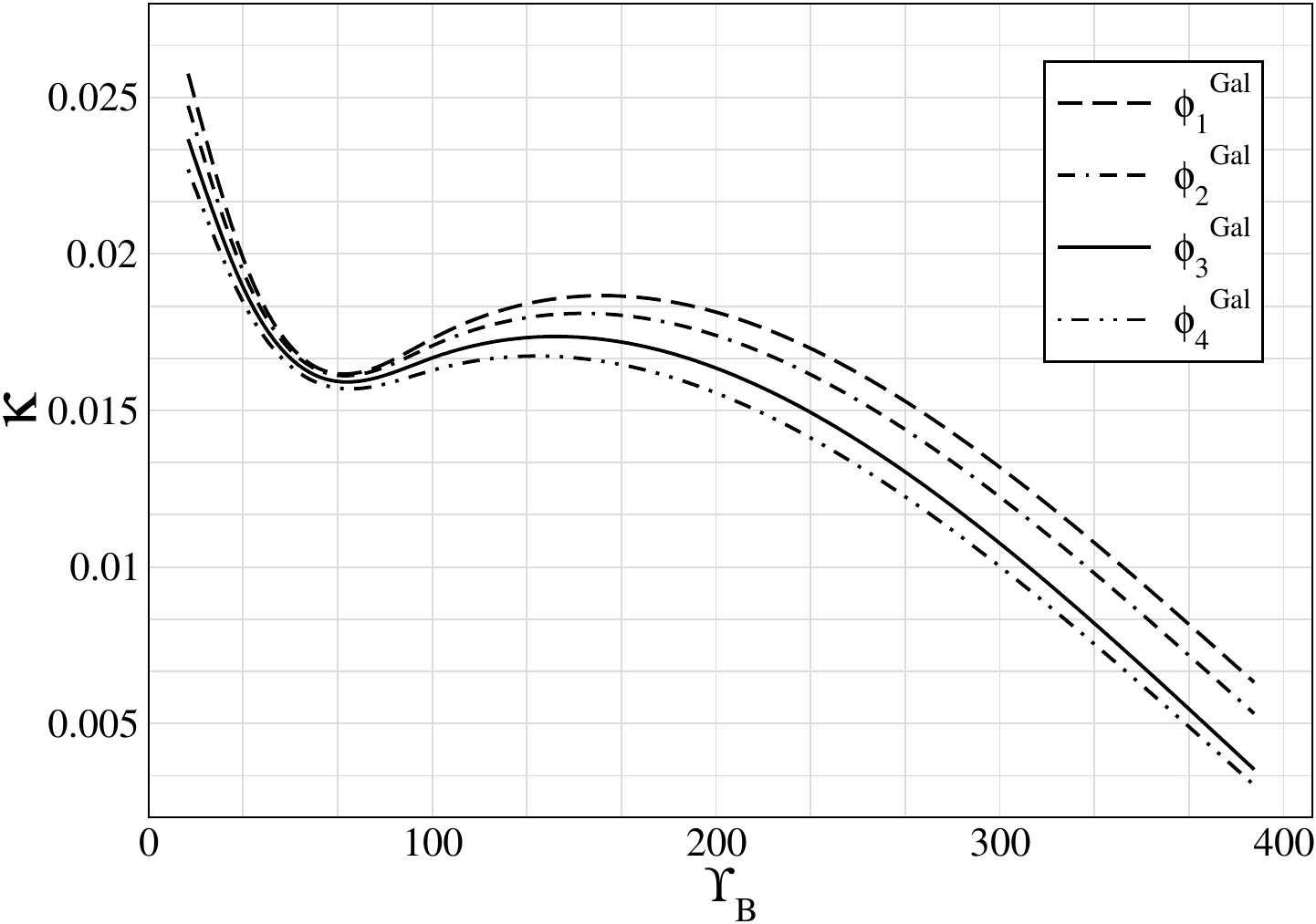}}
  \caption{We plot this figure to show the variation of the acoustic surface gravity ($\kappa$) for the  galactic potential $\Phi^{\mathrm{Gal}}_i$ with four pseudo Schwarzschild potential. The analysis is carried out for a range of galactic parameters: $\Upsilon_{B} = 14$, $33$, $100$, and $390$. The figure illustrates how $\kappa$ responds to changes in the galactic environment, reflecting the combined influence of black hole gravity and the extended galactic potential on the formation and strength of acoustic horizons in axisymmetric transonic accretion. Lower end of $\Upsilon_{B}$ yield almost same response from all the four cases, however increasing $\Upsilon_{B}$ induces its effect on the surface gravity. For small $\Upsilon_{B}$  the galactic potential is too weak to affect the flow, so the surface gravity remains almost the same for all four pseudo-Schwarzschild potentials. But as $\Upsilon_{B}$ increases, the galactic potential becomes strong enough to modify the flow gradients near the sonic point, which in turn changes the acoustic surface gravity. Also the four cases differ because each pseudo-Schwarzschild potential approximates the black hole gravity in a different mathematical form, so the flow responds differently even though the physical black hole is the same.}
\label{VE-k-para}
\end{figure*}
The acoustic horizon, which is similar to a black hole's event horizon and occurs naturally when transonic accretion flows occur, is where the radial flow speed is equal to the local sound speed. The strength of the acoustic horizon and the way linear perturbations in the flow behave are determined by the surface gravity connected to it. In this section, we examine how the galactic background potential affects the acoustic surface gravity in the framework of four pseudo-Schwarzchild potentials. In the figure \ref{VE-k-para} we plot the variation of the acoustic surface gravity ($\kappa$) with the galactic potential $\Phi^{\mathrm{Gal}}_i$ embedded in four pseudo-Schwarzschild potential. The analysis is carried out for a range of galactic parameters $\Upsilon_{B} = 14$, $33$, $100$, and $390$. These parameters represent different strengths of the galactic contribution to the overall gravitational potential.
The picture depicts how $\kappa$ reacts to changes in the galactic environment. It reflects the combined influence of black hole gravity and extended galactic potential on the creation and strength of acoustic horizons in axisymmetric transonic accretion.  The lower end of $\Upsilon_{B}$ produces virtually the same response in all four scenarios; however, increasing $\Upsilon_{B}$ has an influence on surface gravity.  For small $\Upsilon_{B}$, the galactic potential is too weak to alter the flow, therefore surface gravity remains nearly constant for all four pseudo-Schwarzschild potentials.  However, as $\Upsilon_{B}$ increases, the galactic potential modifies the flow gradients around the sonic point, affecting the acoustic surface gravity. Additionally, even if the real black hole is the same, the flow reacts differently in each of the four scenarios because each pseudo-Schwarzschild potential approximates the black hole gravity in a different mathematical form. This figure demonstrates that the galactic environment, through its extended gravitational field, plays a crucial role in modulating the effective surface gravity and the analogue Hawking temperature associated with the accretion flow.

\section{Concluding Remarks}
\label{conclu}
Our comprehensive studies in the Newtonian framework demonstrates the fact that the global configuration and dynamic behavior of low angular momentum, inviscid, axisymmetric accretion processes around a Schwarzschild black hole are considerably impacted by the presence of a galactic environment. We also see that the shock location and shock characteristics are also get modified in the presence of the galactic environment. We adopted two thermodynamic equations of state in the vertical equilibrium (VE) disc model to cross-examine all these aspects while taking into account four distinct pseudo-Schwarzschild potentials. We will go over each of the situations that yielded noteworthy outcomes one at a time below.

\textbf{(1)}~We examined the variation of specific energy of the accretion flow with the position of the critical points and found that the transonic behaviour of the flow depends  on the specific angular momentum and also on the nature of the gravitational potential (Fig \ref{E-r}). The comparison of the galactic potential $\Phi^{\mathrm{Gal}}_i$ with various pseudo Schwarzschild potentials $\Phi^{\mathrm{BH}}_i$ indicates that the incorporation of galactic potential noticeably changes the positioning of critical points, particularly the outer critical point $r_{c3}$. (see Table \ref{tab:critical-points-VE}).

\textbf{(2)}~ We plot the parameter-space diagrams ($\mathcal{E}$--$\lambda$ space) for the four potential considering adiabatic equation of state and observe that inclusion of the galactic potential $\Phi^{\mathrm{Gal}}_i$ causes the wedge-shaped multitransonic region to shrink and move inward toward the black hole (see figure \ref{PMSCHCOVE}). This implies that the galactic potential compresses the range of conditions under which multitransonic accretion can happen and pushes those conditions inward. We also discuss the phase topology for the four pseudo Schwarzschild potential with the respective galactic potential and  see that the later enlarges the outer region of the flow, thereby modifying the location, position and strength of the shocks (see figure \ref{PP1} \& Table \ref{phase-tab}). The shift in shock location indirectly impact the estimation of many physical variables such as accretion rate, as the post shock temperature and density depends on the position of the shock and on the shock strength. Thereby accurately identifying them by considering the appropriate potential form is crucial while estimating the physical parameters from observed data. Further the analysis of the nature and evolution of these critical points (see figure \ref{nature-eigen-CHCOVE}) confirms the occurrence of saddle–center bifurcations.

\textbf{(3)}~ The comparison of shock-permitting regions (see figure \ref{threeshockCHCOVE} \& figure \ref{shock-ISO}) in the $\mathcal{E}$--$\lambda$ parameter space (adiabatic) and the $T$--$\lambda$ (isothermal)  parameter-space for different galactic parameters $\Upsilon_B$ reveals the influence of the galactic environment on shock formation. Increasing $\Upsilon_B$ reduces and shifts the shock-permitting region toward lower angular momentum values, indicating that a stronger galactic potential suppresses shock formation and modifies the outer flow structure. We also see variation of shock strength with different values of $\Upsilon_B$ (see figure \ref{MPT-phi}). With the galactic contribution strengthens, both the Mach number ratio and the post- to pre-shock pressure and temperature ratios evolve systematically. The compression ratio also changes, demonstrating that the thermodynamic response of the flow to shocks depends sensitively on the galactic environment. Investigating the shock location with increasing $\Upsilon_B$ we see shocks tend to form farther from the black hole as $\Upsilon_B$ increases. The outward movement may arise due to the modification in galactic potential's mass distribution. Consequently, the balance between centrifugal and gravitational forces changes, leading to the relocation of the shock front in a manner consistent with the dynamical influence of extended galactic matter.

\textbf{(4)}~Further, we investigate the variation of the acoustic surface gravity $\kappa$ with galactic parameter $\Upsilon_B$ representing the galaxy's mass distribution embedded in four pseudo-Schwarzschild potential ( see figure \ref{VE-k-para}). We see as $\Upsilon_B$ increases, $\kappa$ exhibits a systematic variation that reflects the dilution of the gravitational gradient near the sonic surface. This may indicate that the galactic environment can substantially modify the analogue gravity properties of accretion flows, influencing both the strength and the location of acoustic horizons in realistic galactic contexts.

Our comparative study of the transonic flow dynamics under the multi-component galactic potential where different pseudo-Schwarzschild potential present the central black hole yield interesting results which represents the variation in critical points location, shrinking shock premises along with variation in shock location for different galactic parameters. The behaviour of the acoustic surface gravity with different pseudo-Schwarzschild potential with its associated galactic mass distribution also yields interesting results. Since the mathematical formulation used to reach at the four potential we used, the variation in associated critical points, sonic points and shock locations also produces versatile result worth mentionening. Associating these four potential with the galactic potential yield more differences in those particular aspects.

\section{Acknowledgment}
RS gratefully acknowledges the Department of Physics, Dhruba Chand Halder College. SC acknowledges the ANRF, Government of India, under the National Post-Doctoral Fellowship (NPDF) scheme (Ref. No. PDF/2025/001982). SN likes to acknowledge Dr. Shubhrangshu Ghosh, SRM University, Sikkim, for attracting attention towards these components of galactic potentials available in the literature.

\section*{DATA AVAILABILITY}
Data sharing may not be applicable to this article as it is mostly a theoretical article. No new data are analysed and the numerically created data that support the results are available to the corresponding author.



\bibliographystyle{mnras}
\bibliography{agn-acc} 

\bsp	
\label{lastpage}
\end{document}